\documentclass[lineno]{our_jfm} 

\usepackage{graphicx}
\usepackage{newtxtext}
\usepackage{newtxmath}
\usepackage{natbib}
\usepackage{hyperref}
\usepackage{color}
\usepackage{graphicx}
\usepackage{psfrag}
\usepackage[dvipsnames]{xcolor}
\hypersetup{
    colorlinks = true,
    urlcolor   = blue,
    citecolor  = black,
}
\definecolor{darkgreen}{rgb}{0.0,0.40,0.0}

\newcommand{\RomanNumeralCaps}[1]
\linenumbers
\def\ie{{i.e.}\ }

\def\cf{{cf.}\ }
\def\etal{{\it et al.}~}

\def\bar#1{\overline{#1}}

\def\od{{\rm d}} 
\def\pd{{\partial}} 

\def\plus{{\scriptscriptstyle +}}

\def\c#1#2{\color{#1}{#2}}

\def\D{\mathrm{D}}

\def\P{\mathcal{P}}

\def\new#1{{\color{black}{#1}}}

\title
{Direct numerical simulation of two boundary layers with the same pressure distribution but different surface curvatures}

\author{P. R. Spalart\aff{1}
        \corresp{\email{prspalart@gmail.com}},
        K. E. Jansen\aff{2}
        \and G. N. Coleman\aff{3}}
      
\affiliation{
\aff{1}Woodinville, WA 98072, USA
\aff{2}Ann and H.J. Smead Aerospace Engineering Sciences, University of Colorado Boulder, Boulder, CO 80309, USA
\aff{3}Computational AeroSciences, NASA Langley Research Center, Hampton, VA 23681, USA
}

\pubyear{20??}
\volume{??}
\pagerange{?? -- ??}
\date{?? and in revised form ??}

\begin{document}
\maketitle

\begin{abstract}
A pair of Direct Numerical Simulations is used to investigate curvature and pressure effects. 
One has a Gaussian test bump and a straight opposite wall, while the other has a straight test wall 
and a blowing/suction distribution on an opposite porous boundary, adjusted to produce the same pressure distribution. The calculation of the 
transpiration distribution is made in potential flow, ignoring the boundary layer. This problem of specifying a pressure 
distribution is known to be ill-posed for short waves. We address this issue by considering a pressure distribution that is 
very smooth compared with the distance from wall to opposite boundary. It is also ill-posed once separation occurs.
The pressure distribution of the viscous flow nevertheless ended up very close to the specified one, 
upstream of separation, and comparisons are confined to that region. In the entry region the boundary layers have essentially the same thicknesses
and are well-developed turbulence-wise, 
which is essential for a valid comparison. The focus is on the attached flow in the favorable and adverse gradients.
The convex curvature is strong enough compared with the boundary-layer thickness to make the strain-rate tensor drop to near zero over the top of
the bump. An intense internal layer forms in the favorable gradient, an order of magnitude thinner than the incoming 
boundary layer. The effect of curvature follows expectations: concave curvature moderately raises the skin friction, although without creating G{\"o}rtler 
vortices, and convex curvature reduces it. The pressure gradient still dominates the physics. Common turbulence models unfortunately over-predict the 
skin friction in both flows near its peak, and under-predict the curvature effect even when curvature corrections are included.

\end{abstract}

\section{Introduction}
\label{sec:Intro}
Surface curvature is second to pressure gradient in its ability to modify boundary layers, once they have become fully turbulent. Their interplay challenges our physical understanding and turbulence modelling. Here we treat pressure gradient as the primary influence and curvature as the secondary one, by suppressing the latter and comparing quantities between the two flows, most importantly the skin friction.  This echoes work of \cite{SM73} and \cite{CGAN01}, 
except that they arranged an essentially constant pressure distribution, in order to compare the new flow with the well-known boundary layer on a flat wall without pressure gradient. All three studies demand a precise design of the boundary shape and transpiration opposite to the test wall. They used curved impervious boundaries, whereas we use transpiration along a straight line which is simpler in a simulation with a spectral solver. In all curved-surface
cases, which are of `bump' type, the single stretch of convex curvature is stronger than the two concave ones.  For the curved-surface considered here,  
the convex curvature extends over about 12 boundary-layer thicknesses and $33^\circ$ of turning; this is much less than in the So-Mellor flow, and there is no attempt to reach any asymptotic state in the streamwise direction. It is more typical of the hinges of deflected control surfaces or of leading-edge regions on airplanes, which is of course a strong motivation. The crucial ratio $\delta/R$ of boundary-layer thickness to radius of surface curvature was near 0.08 for them, and peaks at 0.07 for our flow, in which curvature follows a very strong favorable pressure gradient (FPG) and the attendant creation of a sub-boundary-layer (for which of course $\delta/R$ is far smaller).

The curved surface geometry is the longitudinal-centreline cross-section through the midspan of the `Speed Bump' (SB) flow created at Boeing~\citep{jpS19} --
which was proposed to support research addressing the concern, widespread in mechanical and aerospace engineering, 
over the prediction of smooth-body separation by turbulent CFD~\citep{SKADGLM14}.
In the present study, interest stops a little before separation, because the potential-flow control fails. 
Naturally, separation will depend on the history of turbulence physics in the attached boundary layer.
The SB configuration has been the subject of several experiments in its full three-dimensional version \citep{WSSRF20, GGTCLM22},
and of several DNS studies based on the surface geometry along the midspan cross-section and spanwise periodic conditions \citep{BJ21, SSST21, UM22, PBEJ24}.
The aspect ratio of the full shape is high enough for comparisons between 2D and 3D flows to be very instructive although not exact. 
No experimental results will be used here, assuming that the DNS runs are both of high accuracy.

The attempt to create a flow on a straight wall with the SB pressure distribution brought two dangers, one within the inviscid realm and the other in the viscous realm. The first is that the transpiration distribution is calculated within potential-flow theory, and the problem of solving Laplace's equation as it appears in that theory upwards from a line at $y=0$ (namely $\nabla^2\phi=0$ with both $\phi(x,0)$ and $\phi_y(x,0)$ specified) is ill-posed in the sense of Hadamard, because short waves grow exponentially at a rate proportional to their streamwise wavenumber. This will be put into equations below, and such short waves will be excluded. The second is that in viscous flow the desired pressure could have been unreachable due to separation making the flow depart massively from what the potential-flow equations provided. Here, we can report that no large difference is observed upstream of separation. As a result, the classical principle that an attached boundary layer is fully determined by its pressure distribution and its complete state (mean velocity and turbulence) at an upstream station applies. This is key to both the physical insight and the quantitative comparisons with turbulence models.

\def\ReL{\mathrm{Re}_L}

By convention the SB Reynolds number $\ReL$ is based on the width $L$ of the wind tunnel and the upstream reference speed $U_\infty$; 
the full streamwise length of the Gaussian bump is near $0.7L$, while its height is $0.085L$. DNS was first run with $\ReL=10^6$, 
which brings about a partial relaminarisation under the FPG \citep{BJ21}; 
this greatly complicates the assessment of RANS turbulence models. The DNS here are at $\ReL=2\times 10^6$ 
and are free of relaminarisation. Reaching this range for a flow with substantial pressure gradient, 
as opposed to a channel or constant-pressure boundary layer, is on the edge of what is allowed by current computing power.

The paper is as follows. In \S 2 we introduce the simulation on the curved geometry, and then provide details of the formulation over the straight wall. 
In \S 3 the fidelity of the DNS is considered, both in terms of the individual accuracy of the new case (the curved case having already appeared in the archival literature) and the suitability 
of comparing the curved- and flat-surface flows to isolate surface-curvature effects.
Results first confirm the close correspondence of their pressure distributions and inflow boundary-layer states,
and then display the differences in skin friction and other aspects. In \S 4, results of RANS turbulence modelling 
illustrate its substantial failure in both flows, and the limited success of curvature corrections; in a sense, these findings justify the study. Conclusions are in \S 5.

\section{Definition of the simulations on the curved and the straight geometry}

\begin{figure}
\centering
\centerline{
\hbox{
\includegraphics[width=0.80\textwidth]{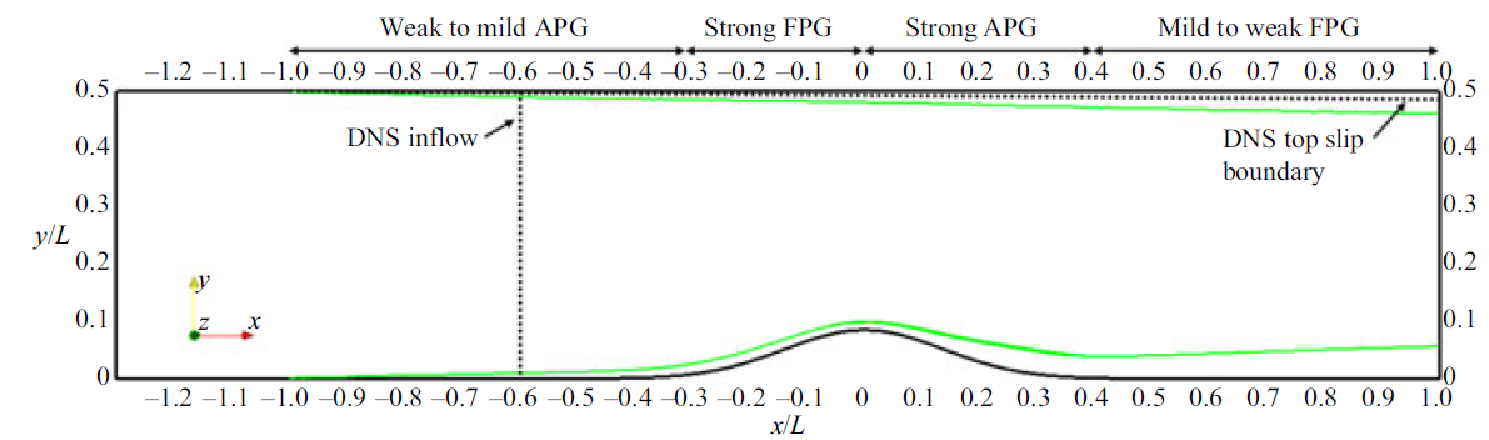}
}
}
\caption{Geometry and boundary-layer thicknesses (green curves, via preliminary RANS solution) for Case~C (figure from \cite{BJ21}).}
\label{fig:Geom}
\end{figure}

\subsection{Case C: Original curved-surface DNS}

Full details of the SB DNS at $\ReL=2\times 10^6$, referred to herein as Case~C, can be found in \citet{PBEJ24};
a brief summary follows. 
For the SB experiment at $2\times 10^6$, the freestream Mach number was only $M_\infty=0.09$ and thus the 
incompressible flow equations where discretized with a stabilized finite-element method \citep{WJ99} 
and implicit time integration \citep{JWH00}.
The DNS used an unstructured grid, with a total of 4.3 billion equivalent grid points,
designed such that the size of each element was less than three local Kolmogorov length scales.  
In local wall units, this corresponds to a maximum of $15$ and $6$, respectively, for the streamwise and spanwise spacing, 
and maximum of $0.3$ for the wall-normal spacing of the wall-adjacent cells~\citep[see][]{PBEJ24}. 
The no-slip boundary condition on the bump surface was combined with convective outflow,  
slip flow at the top ($y=L/2$), and at the inflow a synthetic turbulence generator \citep{SSST14} coupled to a precursor RANS solution.
The streamwise domain extended 
from $x/L=-0.6$ to $1$ (see figure~\ref{fig:Geom}).  Periodic conditions were imposed in the spanwise 
direction, over a width of $0.156L$, corresponding to $7.8$ times the maximum attached boundary-layer thickness (which occurs near $x/L=-0.2$).
This domain accurately captured the separation, reattachment, and recovery regions.
The magnitude of the length- and timescales of the turbulence within these regions, especially the separation bubble, 
demand relatively long integration times to provide adequate statistics; in the domain of interest  for 
this paper, the Case~C statistics cover at least 800 local eddy-turnover times, $U_e/\delta$ 
(where $U_e$ is the edge velocity and $\delta$ the boundary-layer thickness) that occurs at the peak of the adverse pressure gradient on the upstream side of the bump. 

\subsection{Case F: New flat-surface DNS}

\begin{figure}
\centering
\centerline{
\hbox{
\psfrag{x}[][][1.0]{$x/L$}
\psfrag{q}[][][1.0]{$U/U_\infty$}
\psfrag{y}[][][1.0]{$-C_p$}
\psfrag{u}[][][0.45]{$U_{\mathrm{top}}$}
\psfrag{s}[][][0.45]{$U_{\mathrm{slip}}$}
\psfrag{v}[][][0.45]{$V_{\mathrm{top}}$}
\psfrag{C}[l][lc][0.65]{Case C}
\psfrag{F}[l][lc][0.65]{Case F}
\includegraphics[width=0.60\textwidth]{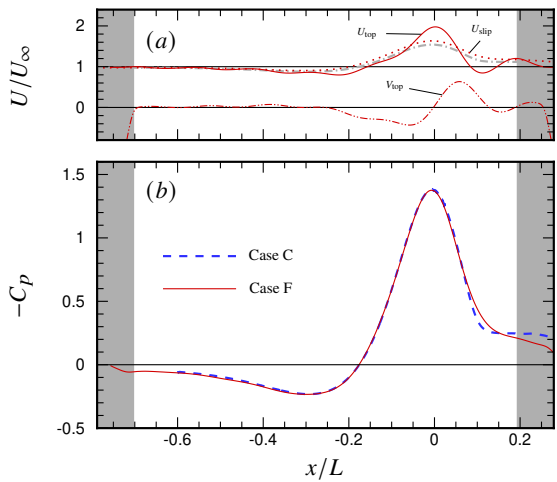}
\put(-182,165) {$(a)$}
\put(-182,109) {$(b)$}
}
}
\caption{Streamwise variation of ($a$) slip velocity at $y=0$ and transpiration at $y=Y_\mathrm{top}$ for Case~F, and 
($b$) pressure at no-slip walls of Cases~C and F.  
Grey dash-dot curve in ($a$) is the slip velocity 
$U_\infty \sqrt{1-C_p}$ given by Case~C wall-pressure distribution.
The grey zones are the fringe regions.
}
\label{fig:C_p}
\end{figure}


Case~F is an incompressible turbulent boundary layer over a flat surface subjected to the Case~C pressure distribution.
The DNS is performed with the fully spectral algorithm described in \cite{SMR91}:  
the velocity field is expanded in terms of Galerkin Fourier/Jacobi divergence-free basis functions, 
such that the variations in the streamwise $x$ and spanwise $z$ directions are periodic, 
while those in the wall-normal direction are over the semi-infinite $0 \le y \le \infty$ domain, with the no-slip wall at $y=0$.
As in \citet{CRS18}, a `fringe' inflow/outflow treatment allows the Fourier method to capture the pressure-gradient-induced 
mean spatial evolution and thickening of the boundary layer, which the fringe terms thin out before it re-enters the active domain.  
Also as in \citet{CRS18}, the desired pressure-gradient profile -- in this case that from Case~C --  
is imposed through an irrotational `transpiration field', as follows. 

We begin with the virtual slip-velocity profile given by the Case~C wall-pressure distribution and Bernoulli's equation 
(shown by the grey/dash-dot curve in figure~\ref{fig:C_p}$a$).  Our objective is to specify the transpiration field between 
the slip velocity at $y=0$ and an arbitrary $y$ location; this velocity field is combined with the `vortical' computational variable to 
produce the full DNS solution with the desired wall-pressure distribution (see \citet{SC97}, 
in which however a simple transpiration distribution was imposed and adjusted to produce the desired extent of flow reversal).  

The transpiration field is defined in terms of the streamwise $U_{\mathrm{top}}$ and wall-normal $V_{\mathrm{top}}$ velocity 
components at a finite distance $y=Y_{\mathrm{top}}$ above the surface.
Because each component of an incompressible, irrotational velocity field satisfies the Laplace equation, 
setting the slip velocity also sets $U_{\mathrm{top}}$ and $V_{\mathrm{top}}$: 
in two-dimensional Fourier space, 
$\od^2 \widehat{u}_i/\od y^2 -k^2 \widehat{u}_i= 0$ 
[$i=(1,2)$, $(\widehat{u}_1,\widehat{u}_2) = (\widehat{u},\widehat{v})$; $k$ is the streamwise wavenumber], 
so that $\widehat{u}_i (k,y) = C_i^+ \mathrm{e}^{+ky} + C_i^- \mathrm{e}^{-ky} = (C_i^+ - C_i^-) \sinh(k\!y) + (C_i^+ + C_i^-) \cosh(k\!y)$.
Invoking the slip-wall boundary conditions 
($\widehat{v}=-\mathrm{i} k \widehat{v} =\od \widehat{u}/\od y = 0$ at $y=0$), reveals 
$U_{\mathrm{top}}$ and $V_{\mathrm{top}}$ via their Fourier coefficients, 
\begin{equation}
\widehat{u}(k,Y_\mathrm{top}) = \widehat{u}(k,0) \cosh( k Y_\mathrm{top}),\quad
\widehat{v}(k,Y_\mathrm{top}) = -\mathrm{i} \widehat{u}(k,0) \sinh( k Y_\mathrm{top}),
\label{eqn:Vtop}
\end{equation}
where $\widehat{u}(k,0)$ is the Fourier transform of $U_\mathrm{slip}$.

\begin{figure}
\centering
\centerline{
\hbox{
\psfrag{x}[][][1.0]{$x/L$}
\psfrag{e}[][][0.9]{$\delta_{_{995}}/L$}
\psfrag{d}[][][0.85]{$\delta^*/L,\theta/L$}
\psfrag{H}[][][0.80]{$H=\delta^*/\theta$}
\psfrag{R}[][][1.0]{$U_\infty \theta / \nu$}
\psfrag{y}[l][lc][0.75]{$y=Y_\mathrm{top}$}
\psfrag{h}[l][lc][0.75]{$H$}
\psfrag{C}[l][lc][0.65]{\c{black}{Case C}}
\psfrag{F}[l][lc][0.65]{\c{black}{Case F}}
\includegraphics[width=0.50\textwidth]{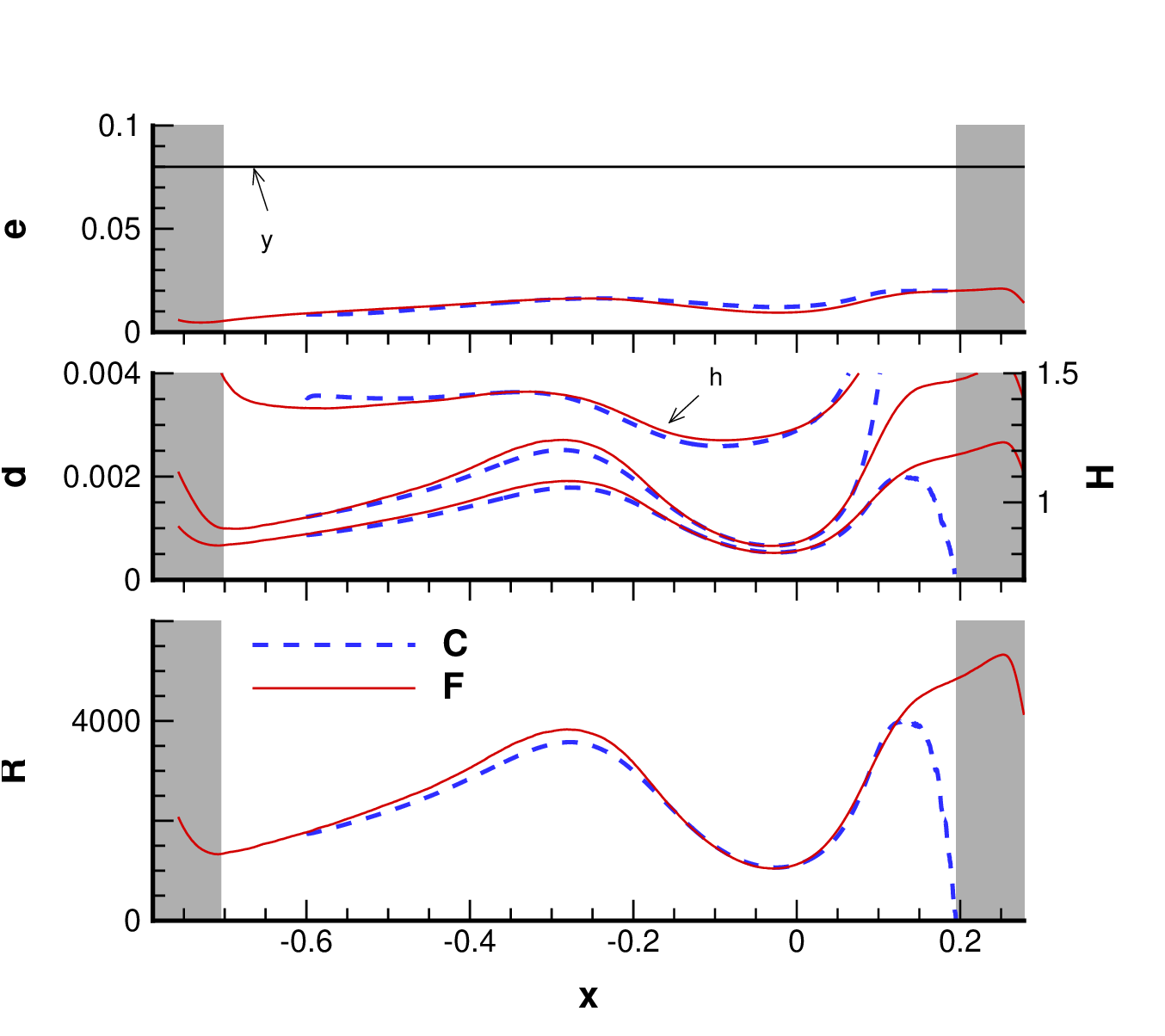}
\put(-150,151) {$(a)$}
\psfrag{x}[][][1.0]{$x/L$}
\psfrag{y}[][][1.0]{$\theta/L$}
\psfrag{c}[l][lc][0.45]{$\c{black}{\int \frac{1}{2} C_f\, \od x}$}
\psfrag{o}[l][lc][0.45]{$\c{black}{(1-C_p)\,\theta}$}
\psfrag{t}[l][lc][0.45]{$\c{black}{-\int \frac{1}{2} \delta^* \,\od C_p}$}
\psfrag{f}[l][lc][0.45]{$\c{black}{-\int(\bar{u'u'} /U_\infty^2)\,\od y}$}
\psfrag{s}[l][lc][0.45]{Sum}
\includegraphics[width=0.50\textwidth]{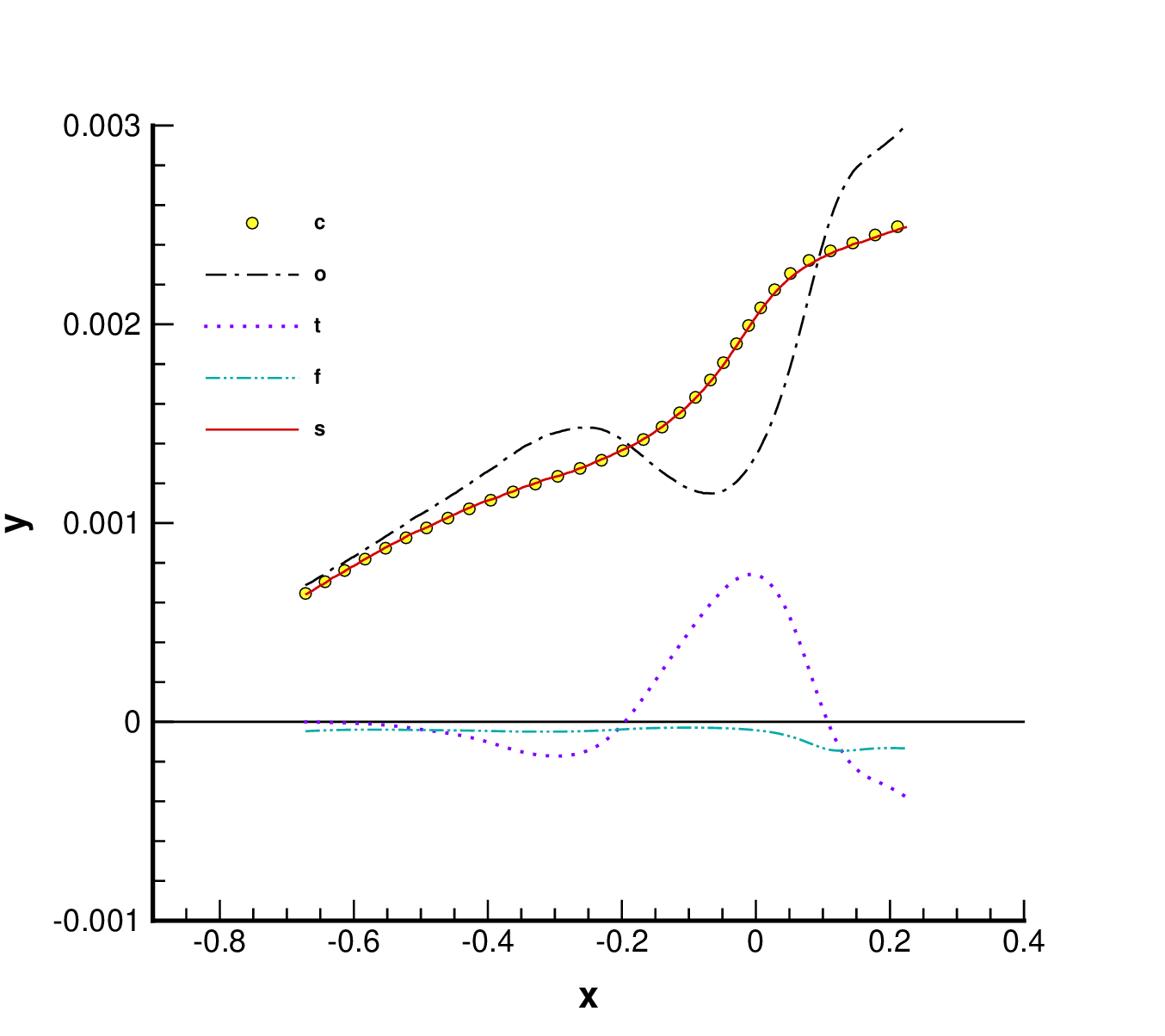}
\put(-155,151) {$(b)$}
}
}
\centerline{
\hbox{
\psfrag{y}[][][1.0]{$y^+$}
\psfrag{U}[][][1.0]{$U^{\,+}$}
\psfrag{C}[l][lc][0.65]{\c{black}{Case C: $x/L=-0.55$}}
\psfrag{F}[l][lc][0.65]{\c{black}{Case F: $x/L=-0.55$}}
\psfrag{c}[l][lc][0.65]{\c{black}{Case C: $x/L=-0.50$}}
\psfrag{f}[l][lc][0.65]{\c{black}{Case F: $x/L=-0.50$}}
\includegraphics[width=0.50\textwidth]{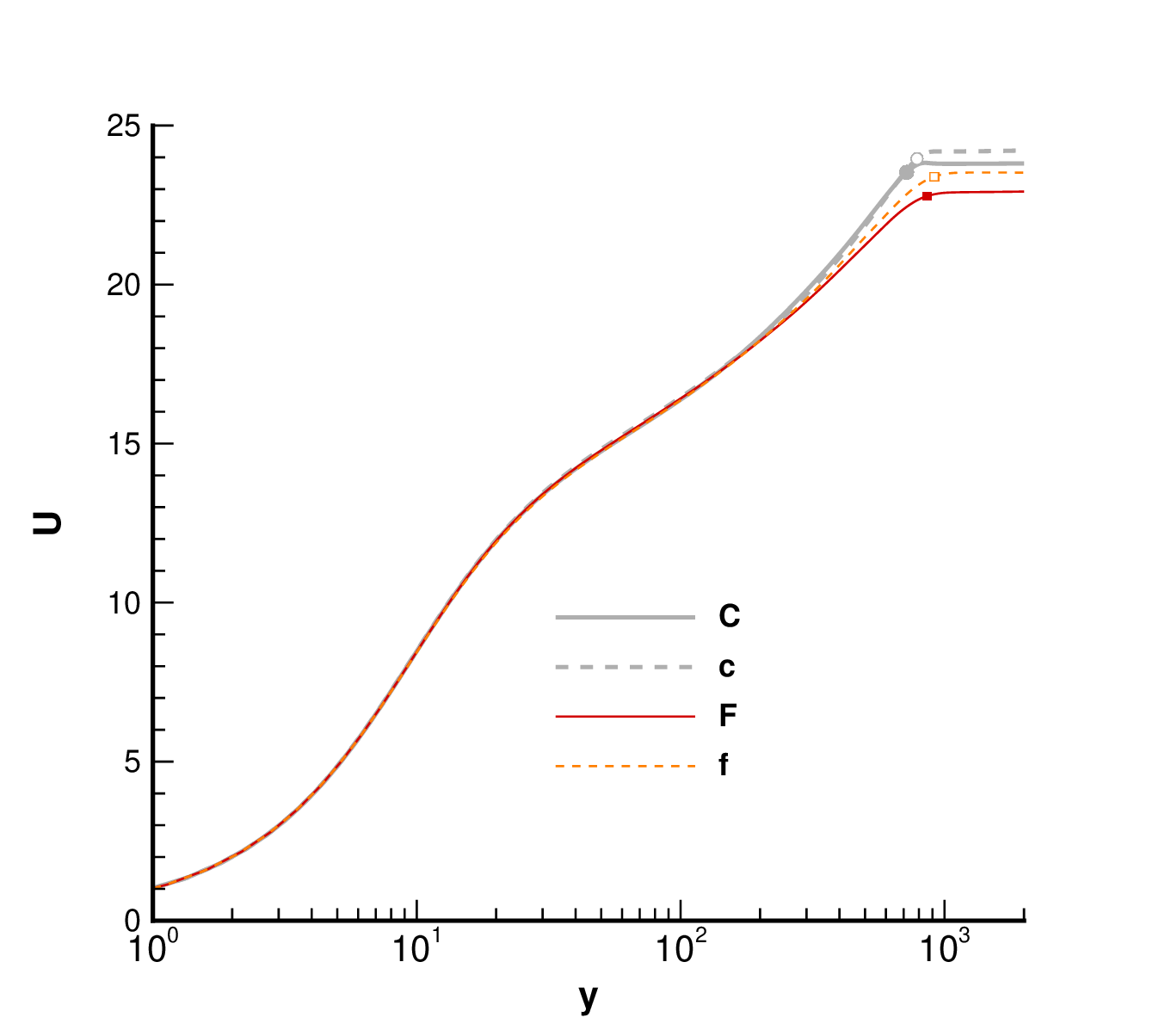}
\put(-155,151) {$(c)$}
\psfrag{y}[][][1.0]{$y^+$}
\psfrag{v}[][][1.0]{$\bar{v' v'}/u_\tau^2$}
\psfrag{U}[][][1.0]{$-\bar{u' v'}/u_\tau^2$}
\psfrag{a}[l][lc][0.45]{\c{black}{$x/L=-0.55$}}
\psfrag{b}[l][lc][0.45]{\c{black}{$x/L=-0.50$}}
\psfrag{c}[l][lc][0.45]{\c{black}{$x/L=-0.45$}}
\psfrag{d}[l][lc][0.45]{\c{black}{$x/L=-0.39$}}
\psfrag{e}[l][lc][0.45]{\c{black}{$x/L=-0.30$}}
\includegraphics[width=0.50\textwidth]{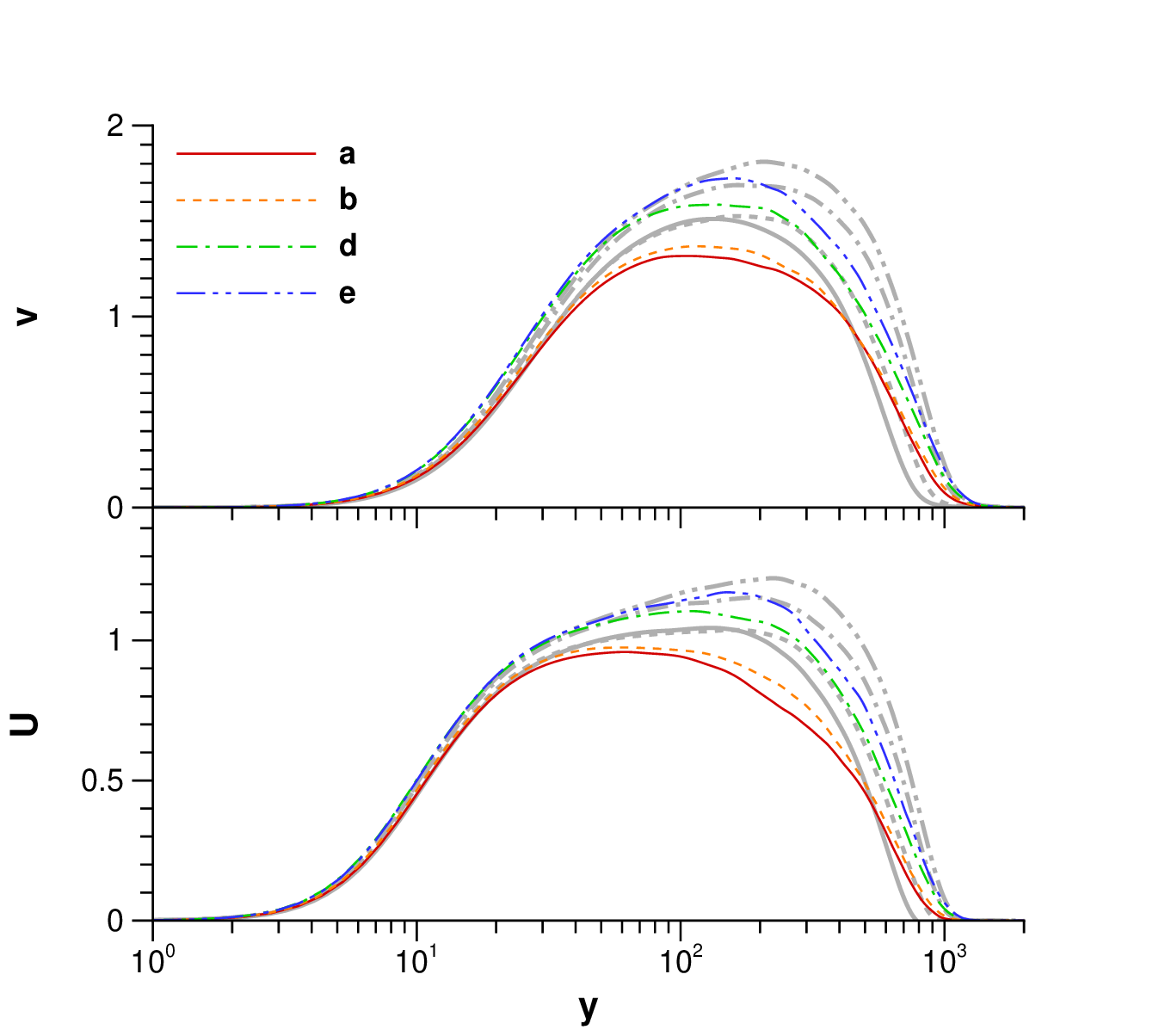}
\put(-155,151) {$(d)$}
}
}
\caption{($a$)~Boundary-layer thicknesses and momentum-thickness Reynolds number for Cases~C and F, 
($b$)~streamwise-integral of momentum balance for Case~F, and
($c$)~mean velocity  and ($d$)~Reynolds-stress profiles from upstream/mild-APG region for Case~C~(shaded/grey) and F~(color).
Thicknesses $\delta_{995}$, $\delta^*$ and $\theta$ based on mean spanwise vorticity profiles 
\cite[see (3.1) -- (3.4) of][]{CRS18}.
The grey zones in ($a$) are the fringe regions used for Case~F. Symbols in ($c$) denote $\delta_{995}$.
Reynolds stresses for Case~C in ($d$) presented in terms of streamline-aligned $(s,n)$ coordinates~\cite[see][]{PBEJ24}.
}
\label{fig:theta}
\end{figure}

Three comments are in order.
\begin{enumerate}
\item Since $\sinh(k\!y)$ and $\cosh(k\!y)$ grow exponentially with $k\!y$, (\ref{eqn:Vtop}) 
places a severe constraint on the maximum wavenumber $k_\mathrm{max}$ needed to precisely represent the wall-slip velocity $U_\mathrm{slip}(x)$, 
and on $Y_\mathrm{top}$.  Consequently, $U_\mathrm{slip}$ must be very smooth, and $Y_\mathrm{top}$ must be as small as possible 
but with a margin above the vortical/boundary-layer region.  Fortunately, the pressure distribution induced by the Gaussian speed-bump geometry for Case~C 
is smooth enough, and the boundary layer that passes over it is thin enough 
(relative to $Y_\mathrm{top}$; see upper panel of figure~\ref{fig:theta}$a$), that both constraints are satisified for the Case~F DNS.
Fourteen nonzero Fourier modes within the period of the domain which includes the fringe, corresponding to $k_\mathrm{max} Y_\mathrm{top} = 6.6$, are sufficient to 
provide a good representation of the Case~C pressure distribution in Case~F (see figure~\ref{fig:C_p}$b$).

\item Obtaining the Case~F wall pressure is an iterative process.
Although the direct, Bernouilli-equation estimate of the slip velocity is fairly accurate -- in terms of the closeness, relative to the Case~C target,  
of the wall pressure it induces in the DNS -- further improvements were obtained `manually' by slightly adjusting 
$U_\mathrm{slip}$\footnote{After the entire $C_p$ profile from Case~C was multiplied by $0.997$, it was further reduced by subtracting 
two small, asymmetric Gaussian `shims' (centred at $x/L=-0.35$ and $0.008$), before the $U_\mathrm{slip}$ calculation.}. 
The result is shown by the dotted red curve in figure~\ref{fig:C_p}$a$ (\cf broken grey curve, which 
traces the unadjusted slip velocity).  Applying (\ref{eqn:Vtop}) to the adjusted $U_\mathrm{slip}$ yields 
the red dash-double-dot ($V_\mathrm{top}$) and solid ($U_\mathrm{top}$) curves in figure~\ref{fig:C_p}$a$.
Note the nascent waviness in the $y=Y_\mathrm{top}$ profiles, associated with the $\mathrm{e}^{+ k_\mathrm{max} Y_\mathrm{top}}$ 
amplification of the highest wavenumbers in the slip velocity (and how that waviness does not compromise the good agreement 
between the Case~C and Case~F wall pressures).

\item While (\ref{eqn:Vtop}) is used in the DNS to specify the transpiration field over the semi-infinite $0\le y \le \infty$ domain, 
such that the desired profile occurs at $y=Y_\mathrm{top}$, it can also be used to create the Case~F flow in a finite domain, with upper boundary 
at a level $y_\mathrm{top}$, which could be somewhat different from $Y_\mathrm{top}$ -- provided the wall-normal gradient of the streamwise 
velocity there satisifies $\pd \bar{u}/\pd y = \od V_\mathrm{top}/\od x$, 
to ensure the flow remains irrotational along $y=y_\mathrm{top}$.  This finite-domain strategy was used for the RANS-model testing presented below, 
with (\ref{eqn:Vtop}) defining Dirichlet conditions for $V(x)$ as well as $U(x)$ at $y=y_\mathrm{top}$, 
with an a posteriori check on the $\pd U / \pd y = \pd V/\pd x$ condition. Strictly speaking, $Y_\mathrm{top}$ does not enter the DNS, and the velocity field 
given by the cosh and sinh would best be called the `Laplace Field.' The quantity $Y_\mathrm{top}$ only enters the study for post-processing, and to provide a 
boundary condition for other methods, in which case the transpiration velocity would be smoothly returned to 0 at the upstream and downstream ends.
What is needed in the DNS is a band near $Y_\mathrm{top}$ where the velocity field has not yet begun strong exponential behaviour, and the vorticity $\pmb{\omega}$ is 
already essentially 0. This is true because the nonlinear term is expressed as $\mathbf{u}\times\pmb{\omega}$.  
\end{enumerate}

\medskip

The DNS domain extends from $x/L=-0.788$ to $0.278$, with the useful region (unaffected by the fringe treatment) 
between $-0.70$ and $0.19$.  The transpiration plane is set at $Y_\mathrm{top}=0.08L$, while the spanwise period is $0.043L$.
The latter is much smaller than the $0.156L$ used for Case~C, but is nevertheless adequate for our present purposes, since 
Case~F does not direct attention to the separation region and the larger lengthscales associated with it.  The 
spanwise period is at least 2.6 times the maximum boundary-layer thickness upstream of the strong APG ($x/L<0$), 
and over 4.3 times the `inflow' thickness at $x/L=-0.6$ (figure~\ref{fig:theta}$a$) -- 
values similar to those which \citet{BJ21} found were sufficient for their SB DNS at $\ReL=1\times 10^6$. 
  
A total of $5.07 \times 10^9$ collocation/quadrature points ($12\,228$ streamwise, $270$ wall-normal, $1536$ spanwise) is used;
in terms of the wall units defined by the maximum skin friction (near $x=0$; see figure~\ref{fig:Cf_DNS}), 
the streamwise and spanwise grid spacings are respectively $\Delta x^+ = 12.3$ and $\Delta z^+ = 4.0$, 
with the tenth wall-normal point at $y^+ = 6.0$.
The maximum CFL number in the domain is set to $1.7$, which yields an average timestep in wall units of $\Delta t^\plus=0.18$.  
These values satisfy the full-resolution critera for the present scheme 
-- for which the shortest wavelength in the (de-aliased) solution is $3\Delta x_i$ --
to produce accurate first- and second-order statistics \citep{SCJ09}.
Statistics were gathered by averaging over 14\,750 independent fields, spanning a period of 110 eddy-turnover times.
Basic first- and second-order statistics are available on the NASA Turbulence Modelling Resource (TMR) website, 
https://turbmodels.larc.nasa.gov.

\section{DNS results}

\subsection{Quality checks for Case~F}

That the primary goal of imposing the Case~C wall-pressure distribution over a flat plate until shortly before separation has been achieved is clear 
from figure~\ref{fig:C_p}.  (At $x/L=-0.6$, $C_p = 0.0636$ for Case~C, and $0.0641$ for Case~F.)
The thicknesses of the incoming boundary layers are also quite close, with the momentum-thickness Reynolds numbers at $x/L=-0.6$ 
within 2\%, $R_\theta = 1734$ and $1769$ for Cases~C and F respectively (figure\ref{fig:theta}$a$).
Improving this aspect would be quite expensive, requiring a new transient simulation of several flow-through times 
to finely adjust the fringe parameters, followed by the collection of data. 
We also note that the intense FPG weakens the memory the boundary layer keeps of its state near $x/L=-0.6$.

The skin friction also agrees well -- after a brief downstream transient during which Case~C recovers from its 
synthetic-turbulence inflow treatment: the Case~C and F values differ by $7$, $3$, and $0.03\%$ at $x/L=-0.6$, $-0.55$ and $-0.5$, respectively (\cf figure~\ref{fig:Cf_DNS}$a$). 
For Case~F, at $x/L=-0.6$ the Clauser and wall-unit pressure-gradient parameters are respectively  
$\beta \equiv (\delta^*/\tau_\mathrm{wall}) \od P_\mathrm{wall} / \od x = +0.058$ 
and $\Pi^+ \equiv [ \od ( \bar{p}_\mathrm{wall}/\rho) /\od x ] [\nu/u^3_\tau] = +0.00056$, reflecting the  
weak-to-mild adverse-pressure-gradient (APG) conditions (\cf figure~\ref{fig:Geom}); 
the shape factor $\delta^*/\theta$ is $1.36$ and the skin friction is $C_f=0.0037$ here.  
(For comparison, the zero-pressure-gradient (ZPG), $\beta$ = $\Pi^+ = 0$ values at $R_\theta = 1769$ are $\delta^+/\theta \approx 1.4$ and $C_f \approx 0.0038$ \citep{deC62}.)

The mean-velocity and Reynolds-stress profiles mirror the skin-friction behaviour,  
in that in the mild-APG zone, the near-wall regions of both flows agree well for $x/L \in [-0.55,-0.3]$ (compare shaded/grey and color curves in figure~\ref{fig:theta}$c,d$).
Although the agreement across the layer improves with downstream distance, differences in the Case~C and F outer layers remain, especially for the Reynolds stresses.
This illustrates the difficulties involved in matching upstream conditions between numerical solutions of turbulent flows, particularly eddy-resolving ones.
Fortunately, the outer-layer differences in the mild-APG region are not large enough to mask those introduced 
by surface curvature or its absence and/or the strong FPG (see figure~\ref{fig:innerlayer}).

The Case~F mass balance is exact, by construction (recall the divergence-free property of the basis functions in the Spalart algorithm).  
The momentum balance for Case~F is shown in figure~\ref{fig:theta}$b$ 
via the momentum-thickness growth, making use of the boundary-layer approximation 
but including the contribution of the streamwise Reynolds stress.  
The agreement between the sum (solid-red curve) of the pressure, displacement-thickness and Reynolds-stress terms (broken curves) 
and the integrated skin friction (symbols) is quite good (the full, extended balance defined in Appendix~C of \cite{CRS18} 
is even better), which reflects a sufficient duration of the initial transient and the collection interval.
We conclude that the quality of the statistics from the present, flat-wall DNS is sufficient to provide a meaningful comparison with the earlier, Case~C results. 
The most common source for a failure to satisfy this test in a DNS is an under-estimation of the period of the transient r{\'e}gime, 
making the flow during the sample interval not close enough to steady.

\subsection{Comparisons of skin friction and other boundary-layer characteristics}

\begin{figure}
\centering
\centerline{
\hbox{
\psfrag{x}[][][1.0]{$x/L$}
\psfrag{c}[][][1.0]{$\c{black}{C_f}$}
\psfrag{C}[l][lc][0.65]{\c{black}{Case C}}
\psfrag{F}[l][lc][0.65]{\c{black}{Case F}}
\includegraphics[width=0.50\textwidth]{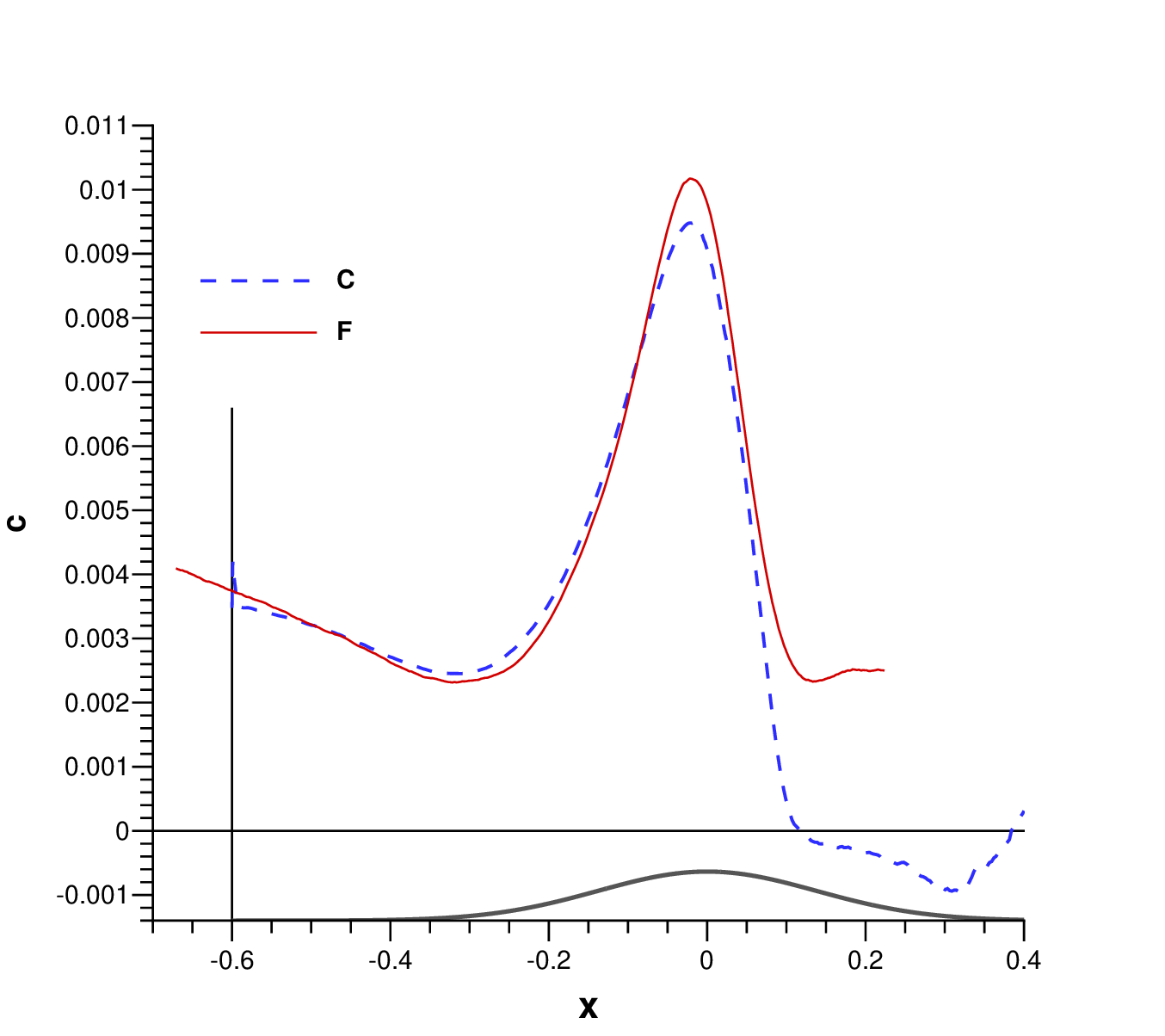}
\put(-155,145) {$(a)$}
\psfrag{c}[][][1.0]{$\c{black}{C_f/(1-C_p)}$}
\includegraphics[width=0.50\textwidth]{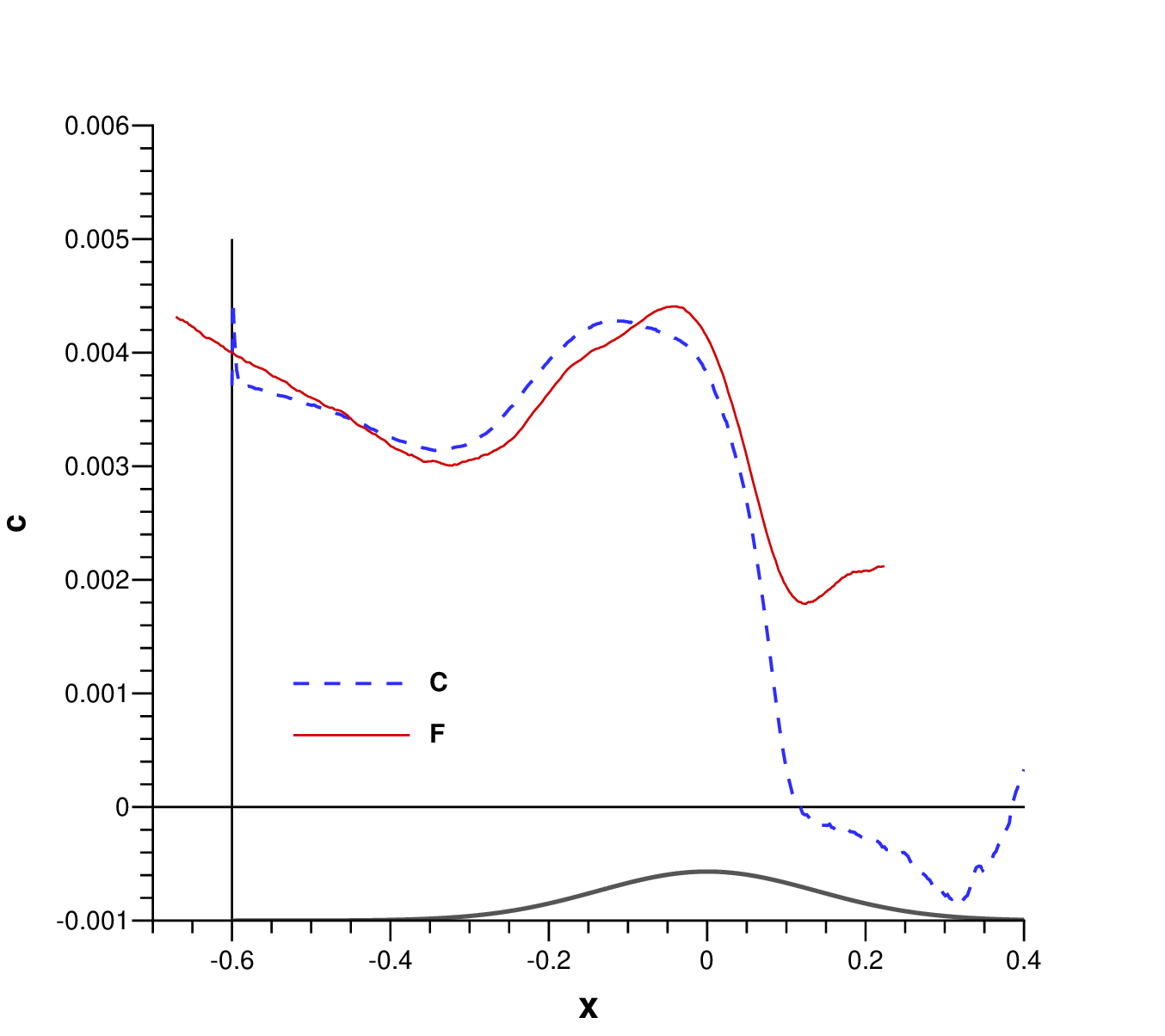}
\put(-155,145) {$(b)$}
}
}
\centerline{
\hbox{
\psfrag{x}[][][1.0]{$U_\infty \theta /\nu$}
\psfrag{c}[][][1.0]{$C_f$}
\psfrag{z}[l][lc][0.65]{ZPG (Coles 1962)}
\psfrag{C}[l][lc][0.65]{Case C}
\psfrag{F}[l][lc][0.65]{Case F}
\psfrag{c}[][][1.0]{$C_f/(1-C_p)$}
\includegraphics[width=0.50\textwidth]{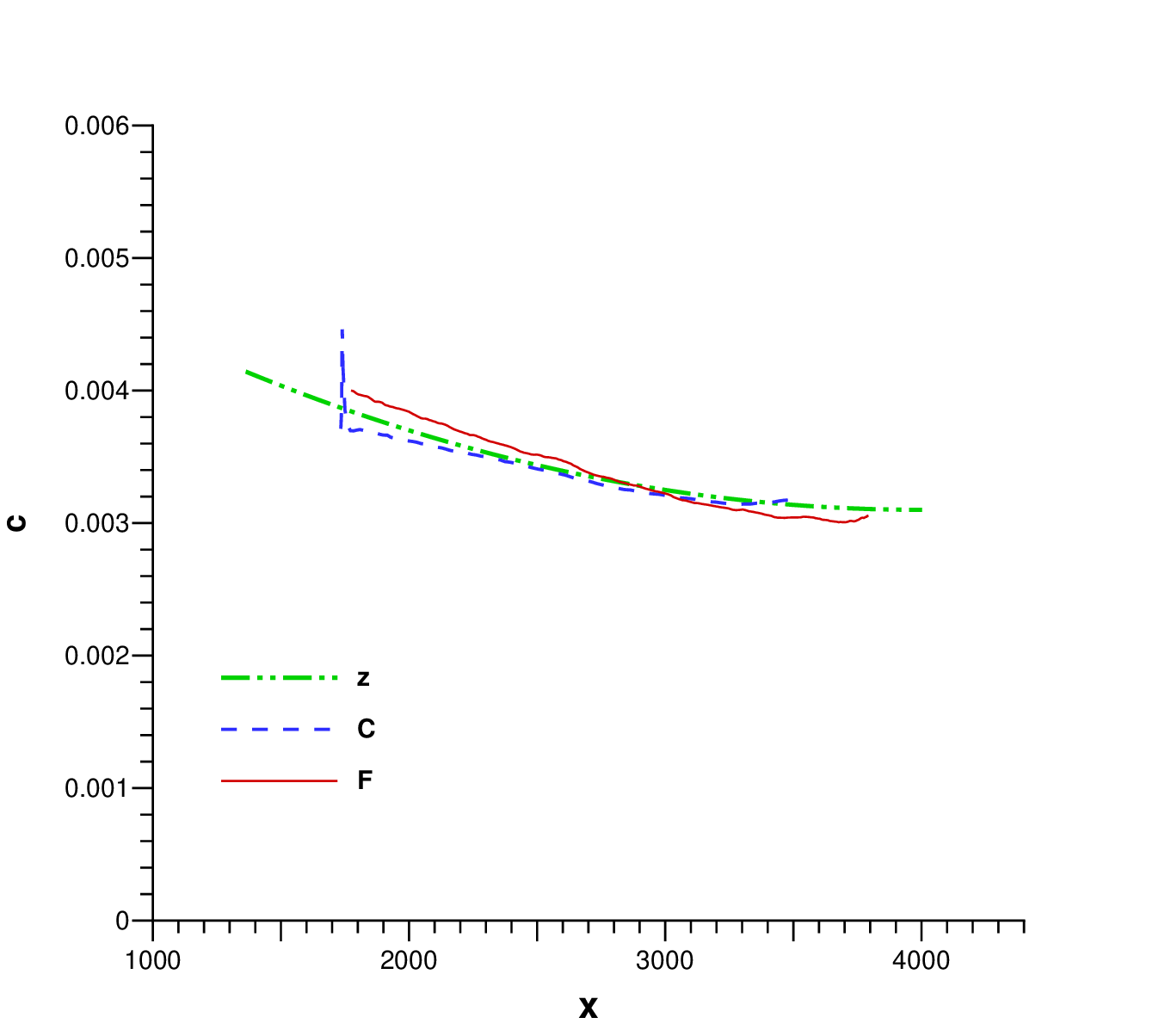}
\put(-155,145) {$(c)$}
}
}
\caption{Skin-friction distributions for Cases~C and F, based on  
($a$)~upstream reference velocity $U_\infty$ and 
($b,c$)~local edge velocity estimate by Bernoulli's equation.
Solid dark-grey curve at bottom of $(a)$ and $(b)$ indicates shape/location of surface geometry for Case~C.
Green dash-dot-dot curve in $(c)$ is interpolant \citep{CRS18} of ZPG boundary-layer data from Coles (1962). 
Reynolds-number range shown in $(c)$ corresponds to $-0.6 \le x/L \le -0.3$.
}
\label{fig:Cf_DNS}
\end{figure}

In figure~\ref{fig:Cf_DNS}, the skin-friction distributions of the two simulations 
are compared using two definitions of the skin-friction coefficient, which contain the same information. 
The first is based on the reference velocity $U_\infty$ upstream of the model, 
and will be called simply $C_f$ as it is the conventional definition: $C_f\equiv \tau_{\mathrm{wall}}/\frac{1}{2}\rho U_\infty^2$. 
The second rests on an approximation of the local edge velocity and dynamic pressure 
of the boundary layer instead: $C_{f\!,\mathrm{loc}}\equiv C_f/(1-C_p)$. 
This comes from Bernoulli's equation, and is limited to attached boundary layers. 
The edge velocity can of course be extracted from the flow field, but only via some conventions regarding the edge of the viscous region, 
and the extrapolation of the inviscid velocity to the wall which is non-trivial due to curvature. The Bernoulli formula ignores these subtleties, but captures the desired effect very well.

Within the attached region, $C_f$ varies by a factor of 5, and the difference at its peak is emphasized. Its peak value of 0.01 would not be 
sustainable by any boundary layer even in FPG. Compare with the sink flow \citep{prS86}, which is the purest example of a boundary layer 
sustaining high $C_f$ values, driven by an FPG; the highest value is about 0.0051. Values as high as 0.0058 have been recorded in boundary layers without 
pressure gradient, but only after vigorous tripping.
This makes intuitive observations difficult. In contrast, $C_{f,\mathrm{loc}}$ varies by only a factor of 2.5, its peak value of $0.0043$ is in the 
typical range, and the visual comparison is much richer. 
The difference in the concave region centred near $x/L=-0.34$ is more visible; differences in the FPG region now appear substantial, whereas for $C_f$ they are hidden 
by the high slope versus $x/L$ (only a figure greatly magnified in $x$ only would achieve this). The $C_{f,\mathrm{loc}}$ curves do not peak at the same $x/L$, whereas those for $C_f$ look 
like one is scaled from the other. The peaks are near $-0.125$ and $-0.05$, respectively; $(1-C_p)$ peaks very close to $x/L=0$, and therefore both the $C_{f,\mathrm{loc}}$ peaks anticipate 
the reversal of the pressure gradient, which is to be expected, but the Case~C flow anticipates more, plausibly because the effect of convex curvature is accumulating 
and suppressing the memory of the concave region, which seems evident near $-0.2$.  A bulge near $x/L=-0.17$, for which we have no physical explanation, is clearly indicated. 
We conclude that the $C_{f,\rm loc}$ quantity is the more useful one for turbulent-boundary-layer studies, and use it hereafter.

\begin{figure}
\centering
\centerline{
\includegraphics[trim={0 2 2 0},clip,width=1.05\textwidth]{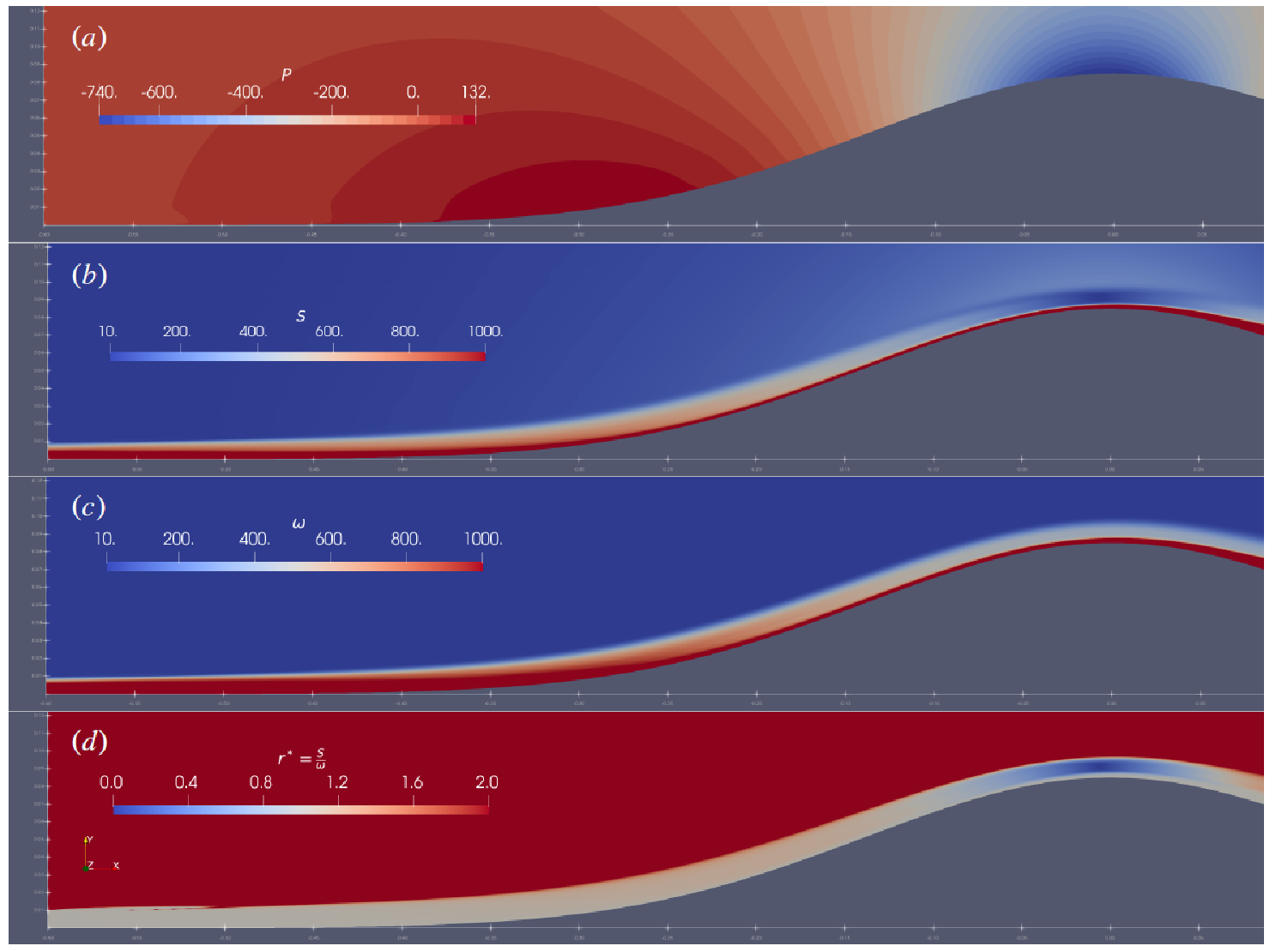}
}
\caption{Case~C contours of mean ($a$)~pressure,
($b$)~strain rate $S$,
($c$)~vorticity magnitude $\omega$, and
($d$)~strain-vorticity ratio $r^*=S/\omega$.}
\label{fig:P(x,y)}
\end{figure}

We also note that Case~F has no flow reversal, in contrast with Case~C, which can be due to flows being more 
sensitive when nearing separation. Reversal in Case~C occurs near $x/L=0.11$, but the agreement on both the pressure and 
skin-friction is still excellent near $x/L=0.07$. Also, Case F has had a lower $C_f$ for some distance, and that 
attached region is the domain of the present study. Another minor point is that the bump inflection point at which
the curvature reverses is near $x/L=-0.14$, but the $C_f$ curves cross only downstream, near $x/L=-0.09$. 
Such a delay or `history effect' is not unusual.

The curvature effect on skin friction pales in comparison with the pressure-gradient effect; this is a genuine finding from the present work. 
In the research flows in the literature, turbulent boundary layers strongly influenced by curvature tend to have large turning angles, such as
$90^\circ$ or $180^\circ$, whereas the sequence here is only essentially `$+15-30+15$' degrees but creates wide $C_p$ variations. 
Technological flows such as airfoils and automobiles of course reach
about 90 degrees of turning at the nose, but with very low $\delta/R$ ratios. Curvature effects are important in some {\em free} shear flows, in particular 
the wake of a wing main element passing over a flap. The wake profiles are visibly asymmetric, and flow reversal in this region can be powerful.
These regions, as well as the shear layers wrapping
around vortices, might deserve a higher priority in modelling than the curved boundary layer does.

\subsection{Evolution of the vorticity and strain rate, growth of the internal layer}

A comparison of the vorticity and strain magnitude in Case~C provides a measure of the departure of the boundary layer from flat-plate behavior
(figure~\ref{fig:P(x,y)}). Simple algebra in a two-dimensional incompressible mean flow leads to $\omega=|\bar{\omega}_z|$ 
and $S\equiv\sqrt{2S_{ij}S_{ij}}=\sqrt{4(\partial U/\partial x)^2+(\omega_z-2\partial V/\partial x)^2}$. 
Over a flat plate, with $\partial/\partial x\approx 0$, $S=\omega$ and this is confirmed in the approach region. 
In the $S$ formula, $(\partial U/\partial x)^2$ is activated in streamwise pressure gradients, while convex curvature is expressed by $\partial V/\partial x<0$, 
which reduces $(\omega_z-2\partial V/\partial x)^2$ (recall that $\omega_z<0$). Over the top of the bump, the pressure gradient crosses $0$ and 
$\partial V/\partial x$ is negative enough to almost nullify $S$ in a significant kidney-shaped region. In contrast, $\omega$ is nearly conserved along streamlines; 
this would be the inviscid behaviour, and the turbulence impact on vorticity is fairly weak over a half-length of the bump, especially as the velocity rose 
much higher than $U_\infty$.  This is fully confirmed by figure~\ref{fig:P(x,y)}. 
The pressure field is shown so as to display its gradient both in the streamwise and wall-normal direction; the latter is what turns the velocity vector.

As mentioned, $\delta/R$ is only 0.07 at the top of the bump, which enters the approximation $\partial V/\partial x\approx-U/R\approx-0.07U/\delta$, but in the outer region $\omega$ is much smaller than $U/\delta$, so that $|\omega_z-2\partial V/\partial x|$ drops to 0, and the quantity inside the absolute value switches sign close to the edge of the boundary layer.

This strong difference between $\omega$ and $S$ is revealing of the intensity of the curvature effect, which has been known in other flows 
to exceed simple estimates often based on $\delta/R$ by an order of magnitude \citep[see][]{pB88} but here we exhibit a kinematic effect, 
whereas Bradshaw was observing changes in the turbulence which may be viewed as consequences of the kinematic effects. 
Note that the ratio $r^*\equiv S/\omega$ is used in many RANS models to render rotation and curvature; 
it is nominally infinite (therefore red in the figure) in the irrotational region, 1 in a simple shear flow and 0 in solid-body rotation 
and on the centreline of a vortex. If injected in the model's source terms, it will steer the eddy viscosity or similar quantities 
with a streamwise delay, which is physically appropriate. With concave curvature, $r^*>1$, giving a qualitatively correct response. 
However, the pitfall in using a single number to reflect curvature is also evident: through the term $(\partial U/\partial x)^2$, 
a pressure gradient of either sign interferes with the depletion of $r^*$; in fact, at the inflection point near $x/L=-0.14$, where 
there is no wall curvature, $r^*$ takes values well in excess of 1 (not visible with the color scale used). This motivated the creation 
of the SA-RC correction, which is complex but more specific than any based on $r^*$ only 
(https://turbmodels.larc.nasa.gov).  
The fact that the sign of the pressure gradient is suppressed by taking the square of $\partial U/\partial x$ 
also makes $r^*$ unpromising as a measure of PG effects.

\begin{figure}
\centering
\centerline{
\hbox{
\psfrag{y}[][][1.0]{$y/\delta_{995}$}
\psfrag{U}[][][1.0]{$\c{black}{U/U_\infty}$}
\psfrag{a}[l][lc][0.5]{\c{black}{$x/L=-0.55$}}
\psfrag{b}[l][lc][0.5]{\c{black}{$x/L=-0.35$}}
\psfrag{c}[l][lc][0.5]{\c{black}{$x/L=-0.30$}}
\psfrag{d}[l][lc][0.5]{\c{black}{$x/L=-0.25$}}
\psfrag{e}[l][lc][0.5]{\c{black}{$x/L=-0.20$}}
\psfrag{f}[l][lc][0.5]{\c{black}{$x/L=-0.15$}}
\psfrag{h}[l][lc][0.5]{\c{black}{$x/L=-0.10$}}
\psfrag{i}[l][lc][0.5]{\c{black}{$x/L=-0.05$}}
\psfrag{j}[l][lc][0.5]{\c{black}{$x/L=0$}}
\includegraphics[width=0.50\textwidth]{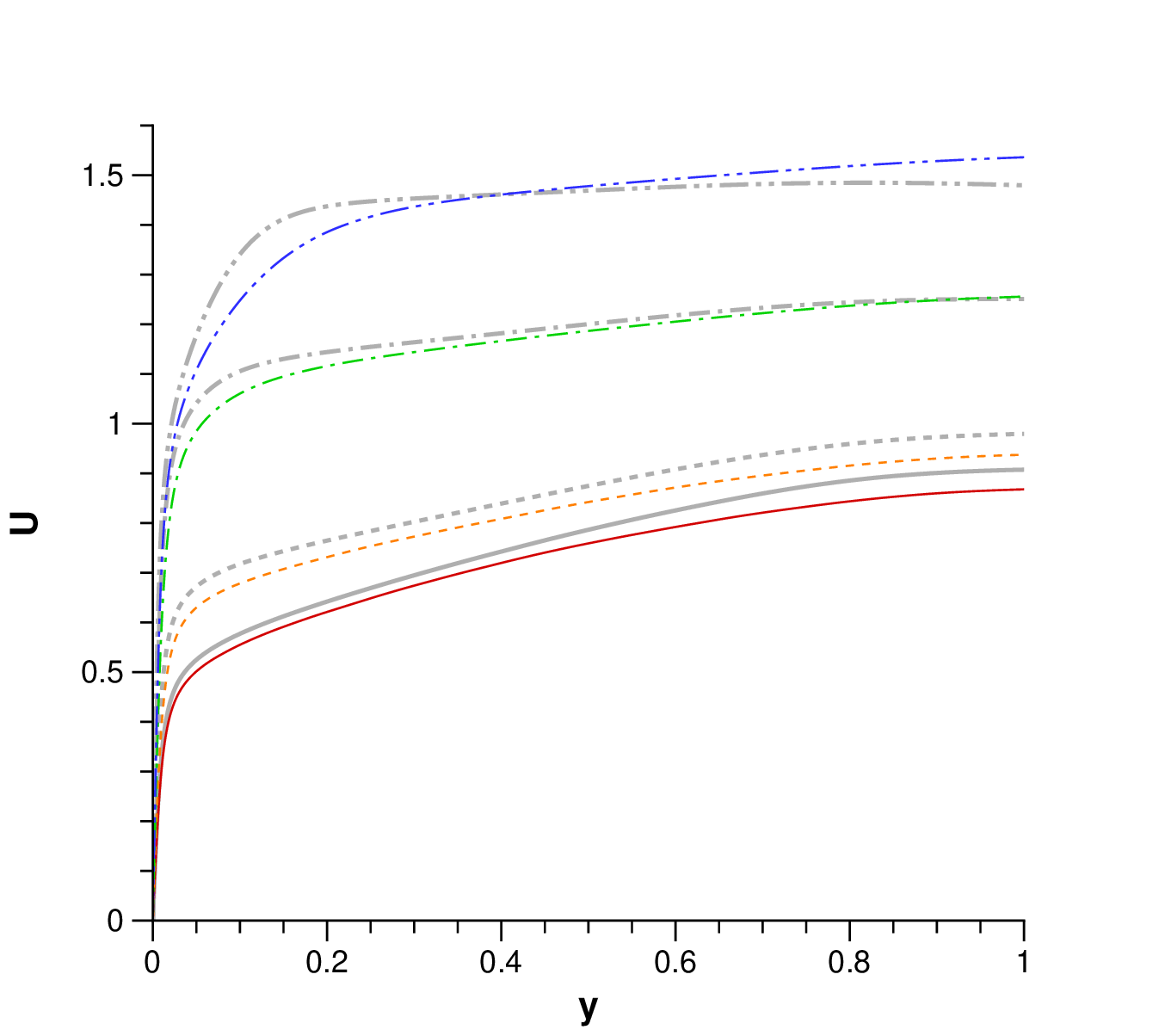}
\put(-160,160) {$(a)$}
\psfrag{y}[][][1.0]{$y^+$}
\psfrag{U}[][][1.0]{$U^{\, +}$}
\includegraphics[width=0.50\textwidth]{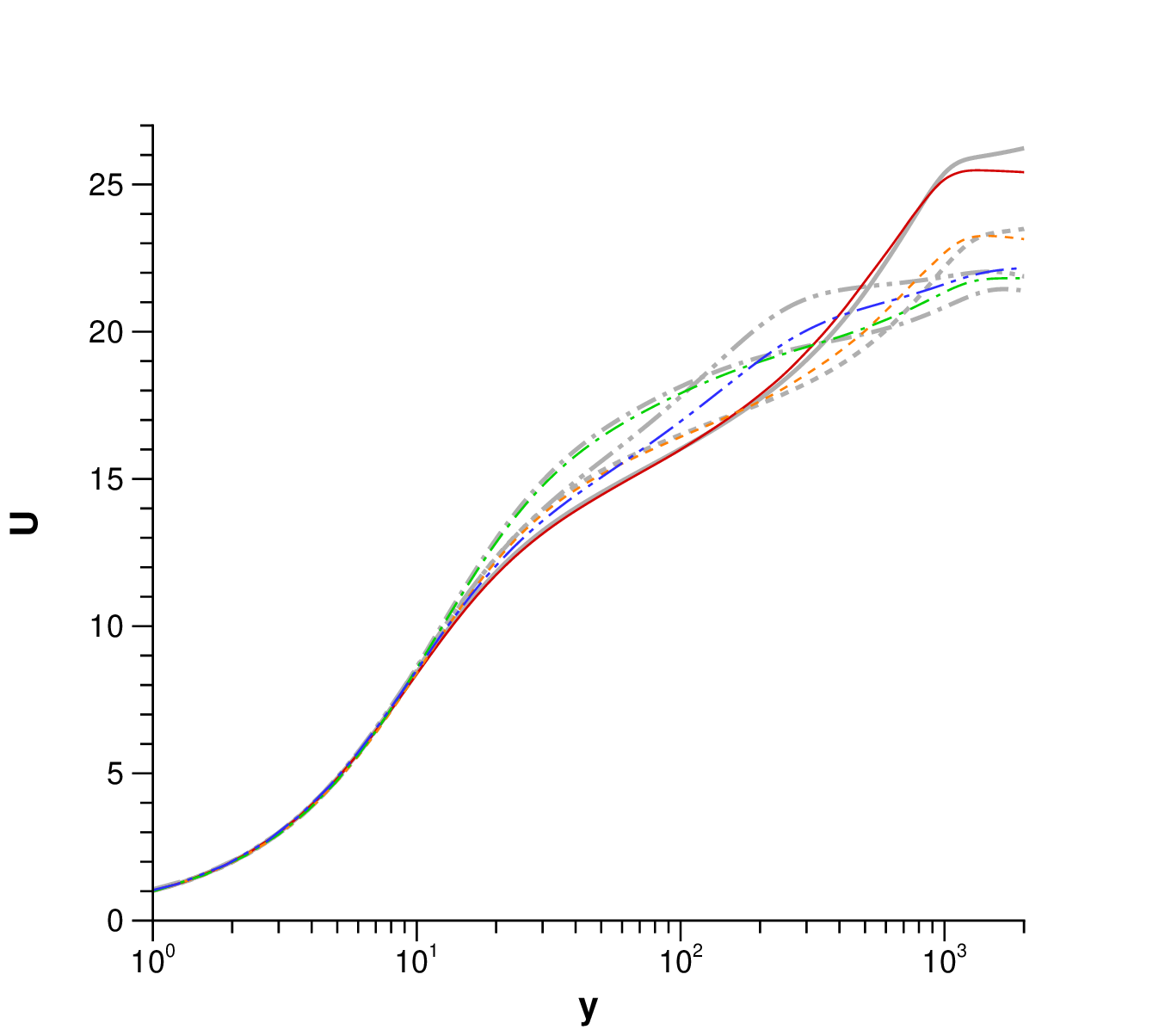}
\put(-160,160) {$(b)$}
}
}
\centerline{
\hbox{
\psfrag{y}[][][1.0]{$y/\delta_{995}$}
\psfrag{U}[][][1.0]{$\bar{v'v'}/U_\infty^2$}
\psfrag{a}[l][lc][0.5]{\c{black}{$x/L=-0.50$}}
\psfrag{b}[l][lc][0.5]{\c{black}{$x/L=-0.35$}}
\psfrag{c}[l][lc][0.5]{\c{black}{$x/L=-0.30$}}
\psfrag{d}[l][lc][0.5]{\c{black}{$x/L=-0.25$}}
\psfrag{e}[l][lc][0.5]{\c{black}{$x/L=-0.20$}}
\psfrag{f}[l][lc][0.5]{\c{black}{$x/L=-0.15$}}
\psfrag{h}[l][lc][0.5]{\c{black}{$x/L=-0.10$}}
\psfrag{i}[l][lc][0.5]{\c{black}{$x/L=-0.05$}}
\psfrag{j}[l][lc][0.5]{\c{black}{$x/L=0$}}
\includegraphics[width=0.50\textwidth]{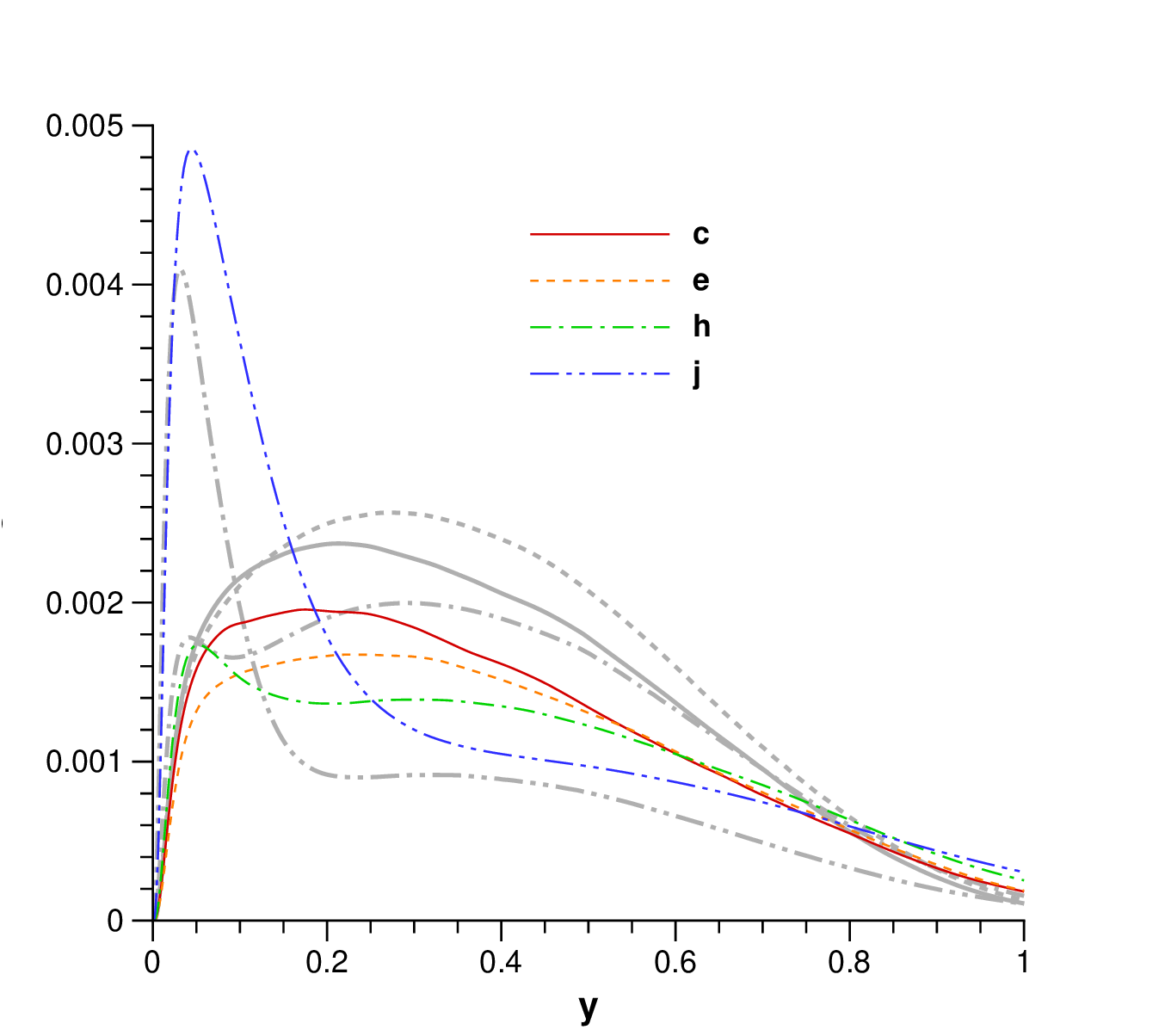}
\put(-160,160) {$(c)$}
\psfrag{y}[][][1.0]{$y^+$}
\psfrag{U}[][][1.0]{$\bar{v'v'}/u_\tau^2$}
\includegraphics[width=0.50\textwidth]{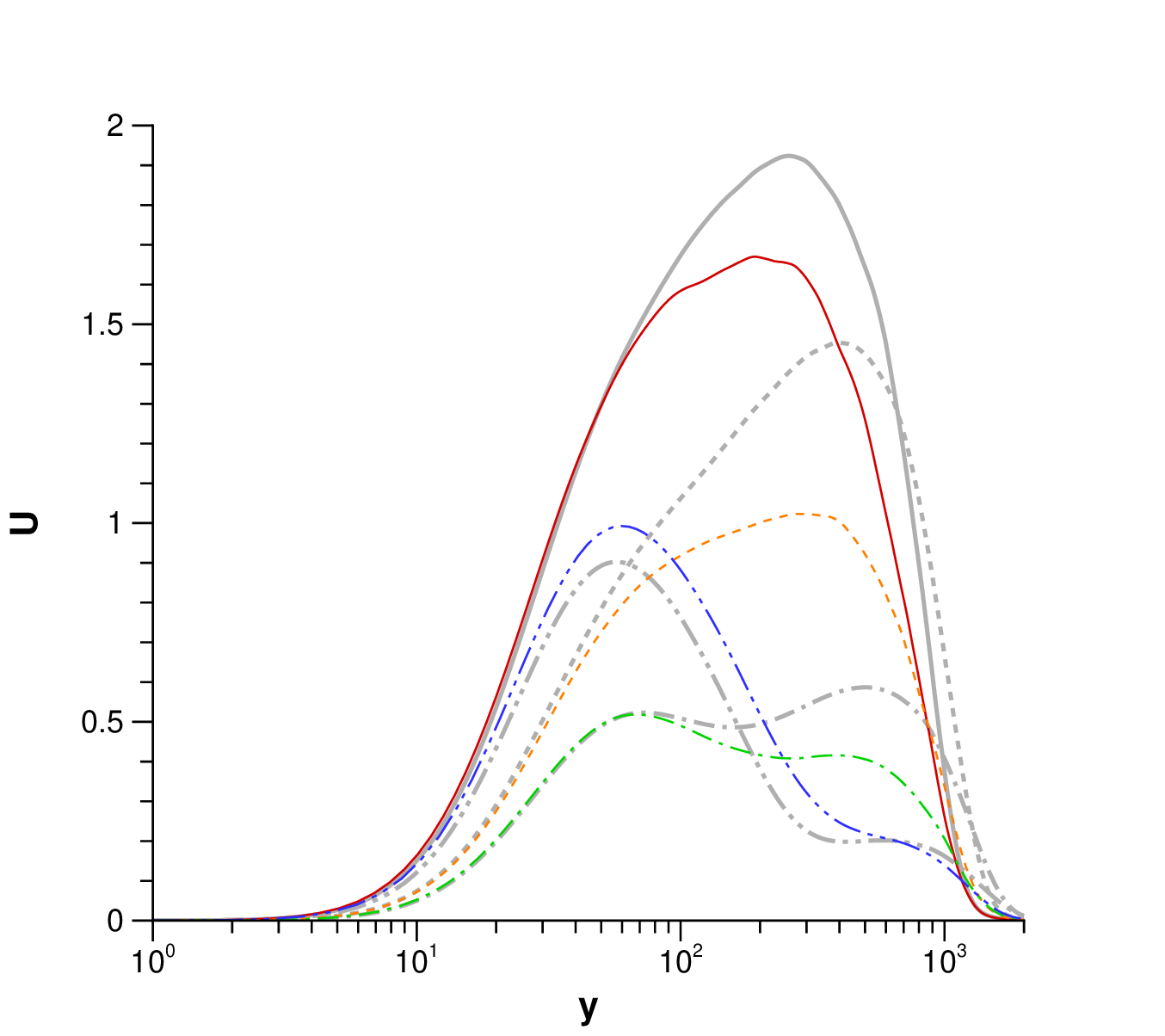}
\put(-160,160) {$(d)$}
}
}
\caption{Profiles of $(a,b)$~mean streamwise velocity and $(c,d)$~wall-normal velocity fluctuations in FPG region, $[-0.3,0]$, 
for Case~C (shaded/grey) and F (color).
Case~C results 
presented in terms of streamline-aligned $(s,n)$ coordinates~\cite[see][]{PBEJ24}.
}
\label{fig:innerlayer}
\end{figure}

Figure~\ref{fig:innerlayer} reveals much about the FPG region. The velocity profiles in figure~\ref{fig:innerlayer}$(a)$ reflect the approximate conservation of vorticity in the outer layer in two ways. 
First, the C case (shaded/grey curves) with convex curvature acquires negative values of $\partial V/\partial x$ in local coordinates, 
leading to less positive values of $\partial U/\partial y$. 
Second, in both cases the reduction of the thickness $\delta$ reduces the velocity difference across the outer layer. The reduction in $\partial U/\partial(y/\delta)$ 
is clear in the graph. This takes place while the velocity difference from wall to freestream is rapidly increasing. This in turn demands a considerable increase in the 
velocity difference carried by the inner region, as is evident in the figure. At first sight, the entire outer region moves upwards as a block. 
An interpretation is that the strong pressure gradient $\od \bar{p}_{\rm wall}/\od x$ injects intense vorticity into the flow, 
as it equals $\nu\partial \bar{\omega}_z/\partial y$. 
Thus, the internal layer which develops in the turbulence has a simple source in mean-flow dynamics. 

Figure~\ref{fig:innerlayer}$(b)$ begins with a normal log layer and a wake marked by the APG at $x/L=-0.3$, and then moves above the Law of the Wall 
to somewhat resemble a laminar profile, with the highest deviation happening at $-0.1$, and the beginning of a recovery at $x/L=0$. This applies to both cases.

In figure~\ref{fig:innerlayer}$(c)$ the apparition of the internal layer on $\overline{v'v'}$ is very sudden, between $x/L = -0.1$ and $0$, 
lagging the rise in velocity difference mentioned above. Curiously, in the outer region $\overline{v'v'}$ first rises, and then almost collapses at $x/L=0$, 
and this even without the stabilizing curvature. This may be attributed to how the direction (eigenvectors) of the stress tensor lag the direction of the strain tensor, 
but the phenomenon is surprisingly strong after a rotation of only about $15^\circ$.

In figure~\ref{fig:innerlayer}$(d)$, $\overline{v'v'}$ now normalized with $u_\tau^2$ falls rapidly. The changes displayed by the curves are very wide, 
regardless of curvature. This makes the failure of RANS models, discussed below, unsurprising. 
An intuitive estimate of the FPG is that at its peak at $x/L=-0.1$, we have $\od\log(U_{\rm slip})/\od (x/L) \approx 2.6$, but the total boundary-layer thickness 
is $\delta/L\approx 0.015$: the product of these two numbers is much smaller than 1. This probably reflects again how boundary-layer turbulence is weak relative 
to pressure forces, even for the present shallow Gaussian bump shape.

\section{RANS results}

\begin{figure}
\centering
\centerline{
\hbox{
\psfrag{x}[][][1.0]{$x/L$}
\psfrag{C}[][][1.0]{$\c{black}{-C_p}$}
\psfrag{d}[l][lc][0.65]{\c{black}{DNS}}
\psfrag{c}[l][lc][0.65]{\fcolorbox{white}{white}{Case C DNS}}
\psfrag{M}[l][lc][0.65]{\c{black}{SST}}
\psfrag{S}[l][lc][0.65]{\c{black}{SA-LRe}}
\psfrag{r}[l][lc][0.65]{\c{black}{SA-RC-LRe}}
\includegraphics[width=0.50\textwidth]{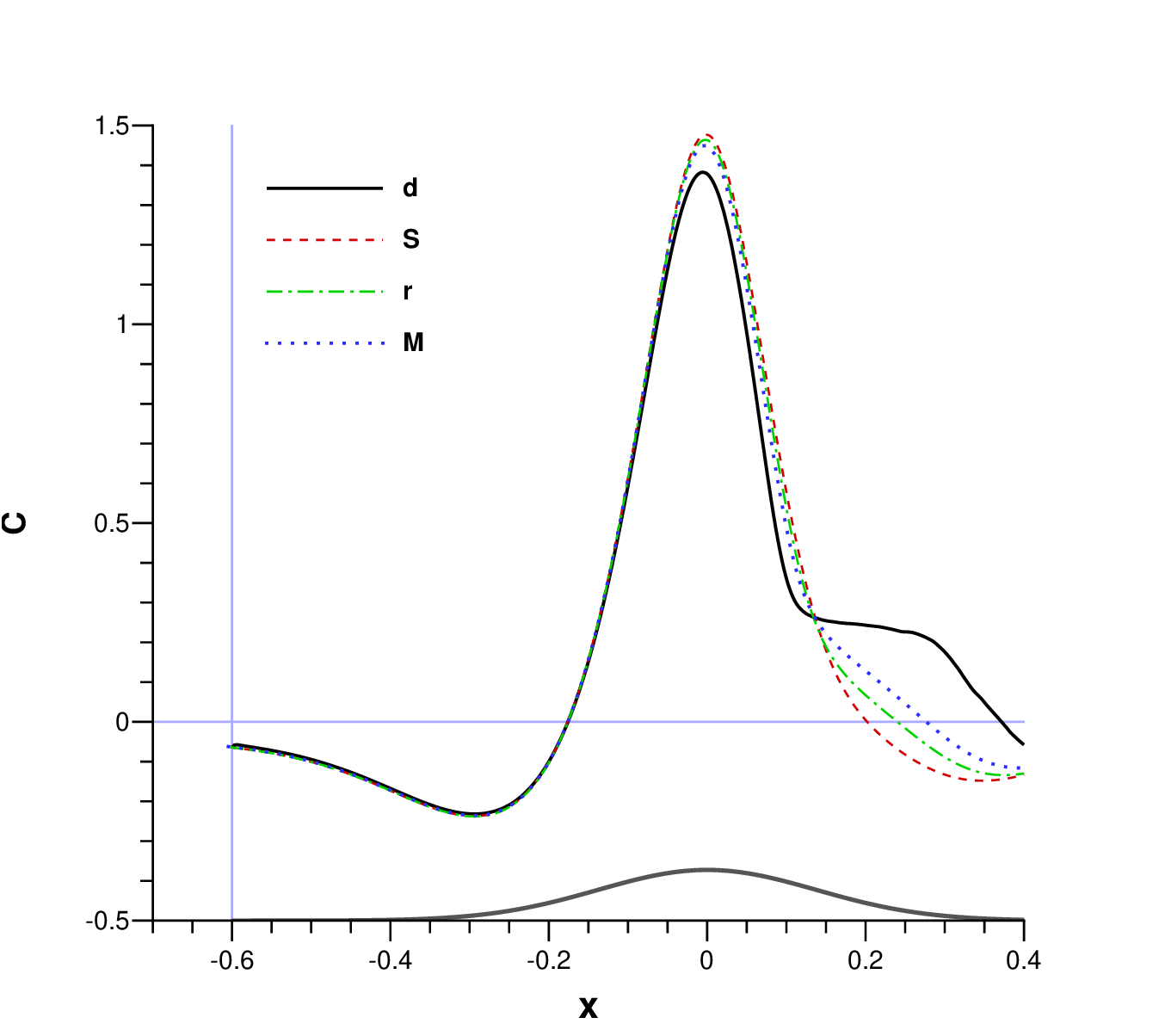}
\put(-160,160) {$(a)$ Case C}
\psfrag{C}[][]{}
\includegraphics[width=0.50\textwidth]{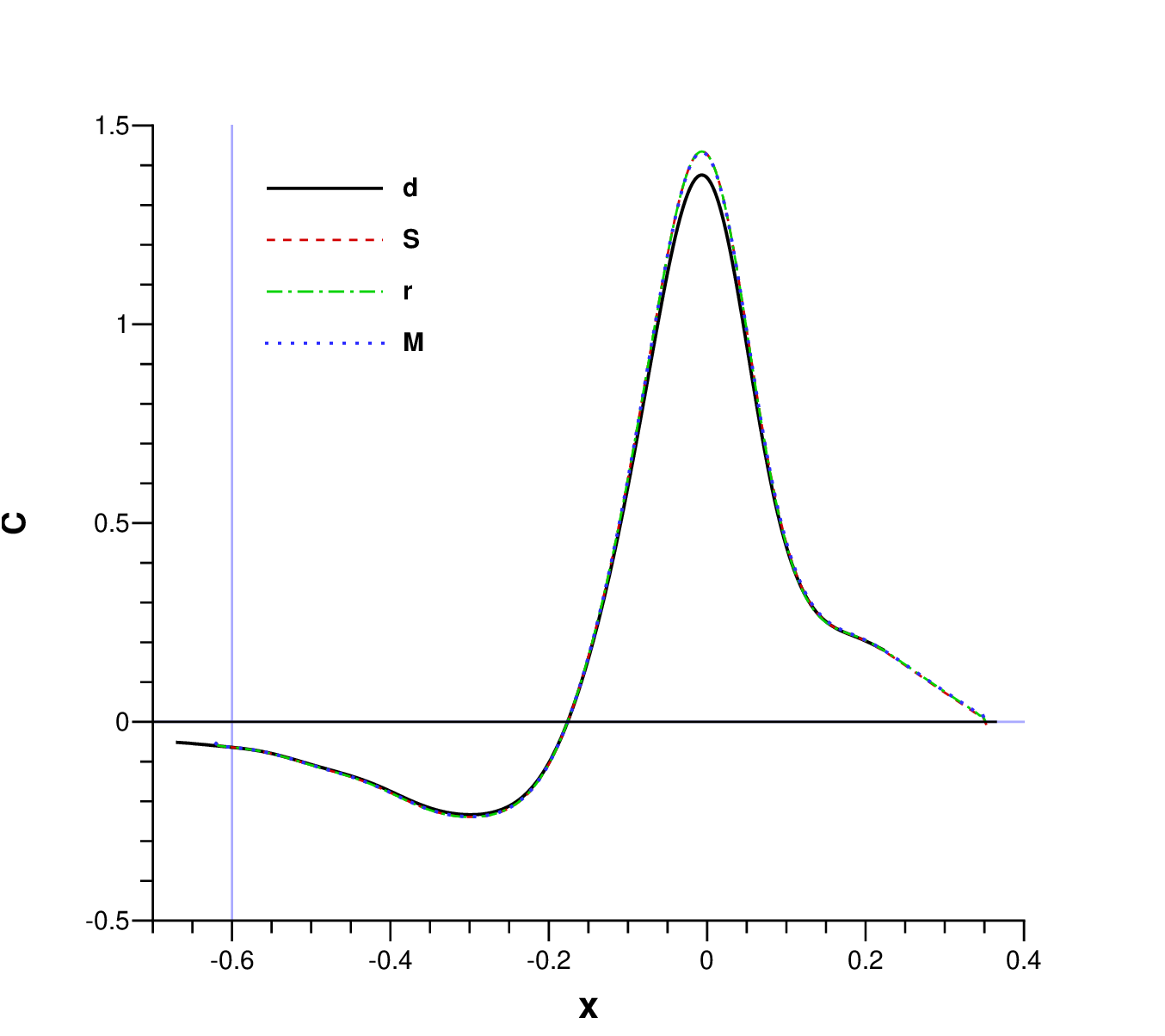}
\put(-160,160) {$(b)$ Case F}
}
}
\centerline{
\hbox{
\psfrag{x}[][][1.0]{$x/L$}
\psfrag{C}[][][1.0]{$\c{black}{C_f/(1-C_p)}$}
\psfrag{d}[l][lc][0.65]{\c{black}{DNS}}
\psfrag{c}[l][lc][0.65]{\fcolorbox{white}{white}{Case C DNS}}
\psfrag{M}[l][lc][0.65]{\c{black}{SST}}
\psfrag{S}[l][lc][0.65]{\c{black}{SA-LRe}}
\psfrag{r}[l][lc][0.65]{\c{black}{SA-RC-LRe}}
\includegraphics[width=0.50\textwidth]{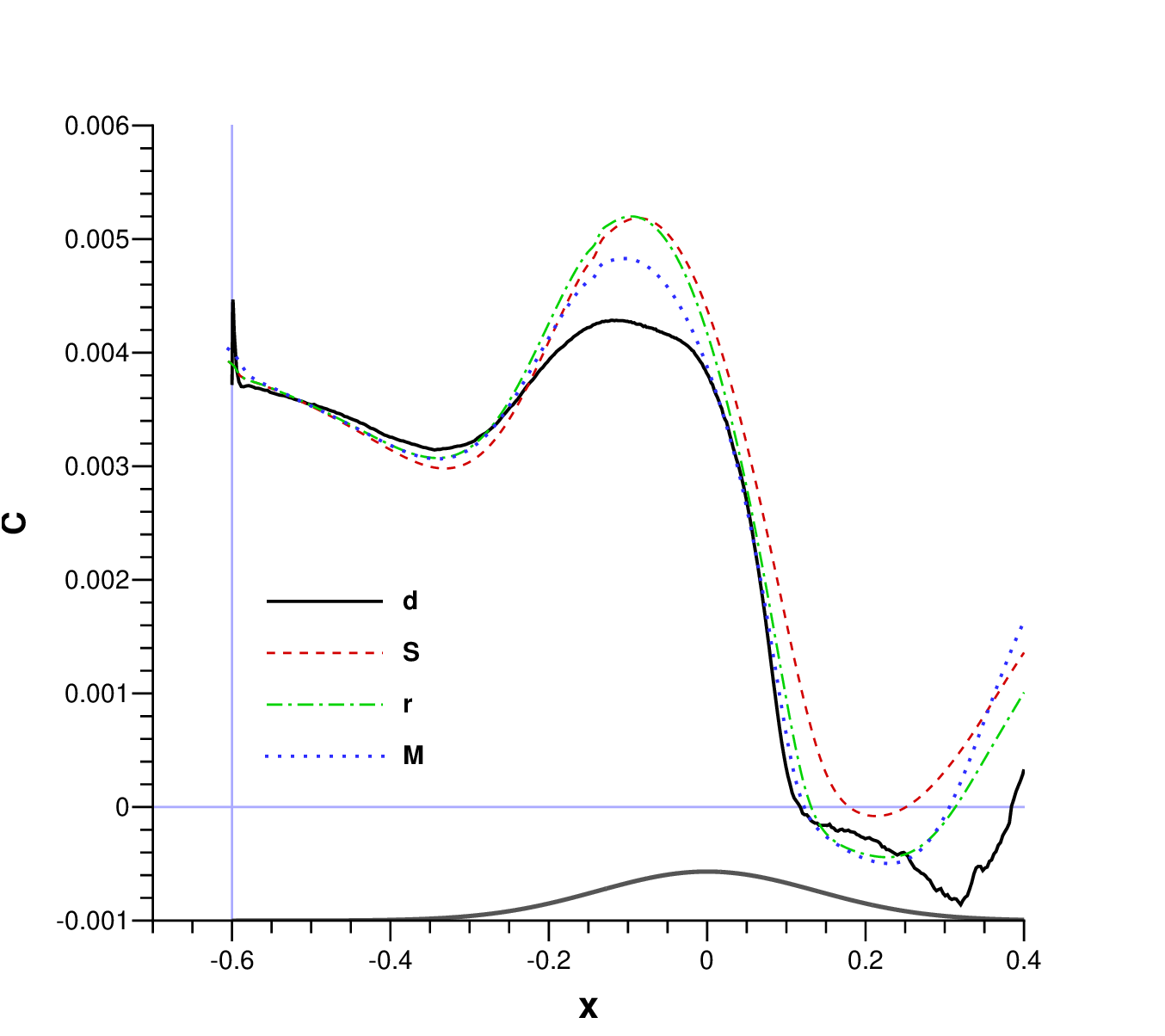}
\put(-160,160) {$(c)$ Case C}
\psfrag{C}[][]{}
\includegraphics[width=0.50\textwidth]{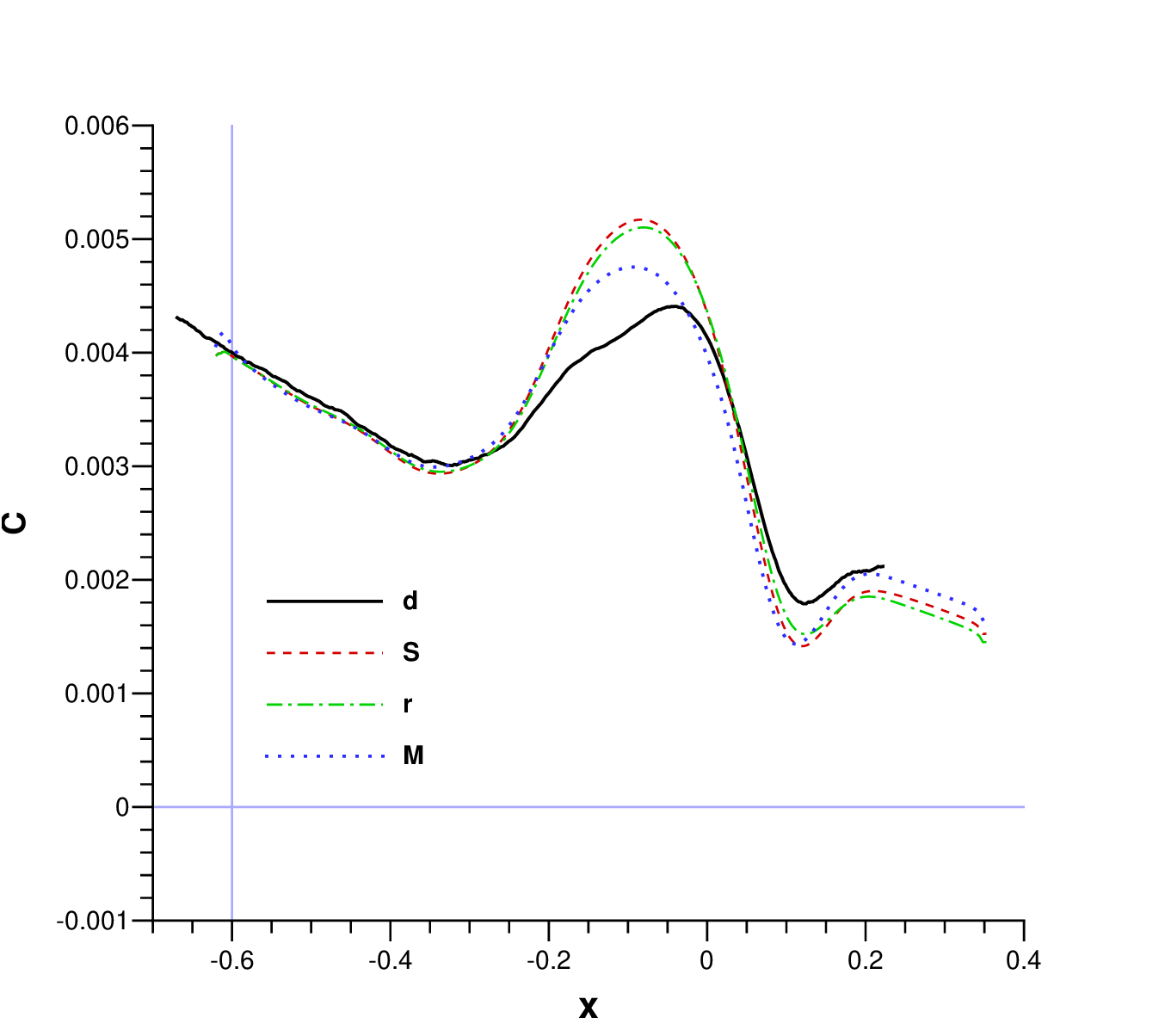}
\put(-160,160) {$(d)$ Case F}
}
}
\caption{Distributions of $(a,b)$~wall pressure and $(c,d)$~skin friction referenced to local edge velocity estimate by Bernoulli's equation, 
for $(a,c)$~Case~C and $(b,d)$~Case~F.  RANS models include SA-LRe~\citep{SG20}, SA-RC-LRe~\citep{SSTS00} and SST~\citep{frM94}. 
Dark grey curve at bottom of $(a)$ and $(c)$ indicates shape/location of surface geometry for Case~C.
}
\label{fig:Cf_RANS}
\end{figure}

The RANS-model tests for Cases~C and F were made using FUN3D \citep{BCDGHJJKLNPRTTWWW20},
with inflow boundary conditions (at $x/L=-0.605$ for Case~C, and $-0.62$ for Case~F) 
defined by the mean profiles at the same station from the corresponding DNS. 
(For example, when applying the SST model, its $k$ variable was taken from the DNS with $\omega$ prescribed such that $k/\omega$ 
closely approximates the eddy viscosity implied by the Reynolds stress and mean-velocity gradient 
at the corresponding $x/L$ station in the DNS. For all the models,  
use of the DNS mean velocities ensured good agreement of the RANS skin friction with the DNS targets at $x/L=-0.6$; 
see figure~\ref{fig:Cf_RANS}.) For the Case~F RANS solutions, the transpiration 
conditions prescribed at the domain top $y_\mathrm{top}$ (defined as $y/L=0.043$) 
were given by the mean streamwise and wall-normal velocities at the same wall-normal location from the Case~F DNS. 
(An {\it a posteriori\/} check was made that the flow at $y/L=0.043$ was essentially irrotational.) 
The streamwise domain, $x/L\in[-0.63,0.35]$, was discretized by $721$ grid points, while $513$ points were employed over the $y/L\in[0,0.043]$ domain.
For Case~C, a no-slip top wall was imposed at $y/L=0.5$ (in contrast to the slip wall used for the DNS; \cf figure~\ref{fig:Geom}),
and the streamwise pressure-gradient/curvature distribution provided by the 
Gaussian SB geometry, centred at $x/L=0$, of the lower no-slip surface, with $513$ points between the two walls.
The $442$-point streamwise domain extended from $x/L = -0.6$ to $1$.

The findings are fairly complex, as displayed in figure~\ref{fig:Cf_RANS}. Again, we believe the $C_{f,\mathrm{loc}}$ presentation 
is much more revealing than that from $C_f$. We do not expand on findings around separation, except to note that all RANS 
models agree with DNS in terms of the skin-friction reversal seen in Case~C disappearing in Case F. The deviations upstream of 
separation are larger than expected, and can be associated with both pressure gradient and curvature.

First examining the mild-APG region, $x/L\in[-0.6,-0.3]$, the models do reasonably well, especially in Case~F, which suggests a full understanding of both DNS and RANS.
The divergence beginning near $x/L=-0.5$ in Case~C appears, then, to be a concave-curvature effect, and one that the SA-RC correction 
only captures to about 40\%. 
We concede that some of the RANS-DNS disagreement here, for both cases, 
could be due to imperfection in the DNS inflow turbulence treatments 
(synthetic turbulence generator for Case~C, fringe zones for Case~F). 
On the other hand, for Case~C, all indications are that any spurious effects of the inflow treatment occur well upstream of $x/L=-0.5$ \citep{PBEJ24}.    
Nevertheless, there is a slight uncertainty here, which again could be explored only through very costly new simulations.

In the FPG, the models rapidly deviate upwards from DNS, in both cases and particularly for the SA model. While separation is the 
major concern of the engineer, we see that the models encounter trouble well inside the attached region, which is often viewed 
as `easy.' Their excess skin friction could thicken the boundary layer and promote early separation, but for SST and SA-RC this does not happen 
(the reversal region is also shorter than in the DNS, in contrast to well-known flows such as the NASA Hump). Curiously, the models do not seem to suffer 
from their history near $x/L=0$, even though in an APG memory effects are enhanced in the von-Karman momentum equation. In the $[0.05,0.15]$ region 
the accumulating effect of SA-RC over the convex wall is beneficial. For Case~F, the impact of the SA-RC model is quite weak, as could be expected.

Around the worst failure of SA, at $x/L\approx -0.1$, it was very reasonable to expect a favorable downward effect for SA-RC, and this is not happening. 
A fair conjecture is that the skin friction there is dominated by the internal layer, and the ratio of its thickness to the radius of curvature is in the 0.007 
region so that one can argue that it has `insignificant' curvature. The RC term is active in the outer part of the boundary layer, 
as illustrated by $r^*$ in figure~\ref{fig:P(x,y)}, but the turbulence there is much weaker than the skin friction and internal layer, 
so that the RC effect does not register at the wall until roughly $x/L=0.05$, as mentioned above.

A general observation is that the models completely miss the shape of the $C_{f,\mathrm{loc}}$ distributions near the peak, 
both returning simple shapes close to inverted parabolas with the peak at $x/L\approx -0.08$ in both flows. It is remotely possible 
that the Reynolds number is still too low to completely eliminate any tendency towards relaminarisation but raising it convincingly, 
such as by a factor 3/2, is not possible presently.

We conclude that the principal challenge is the intense FPG and internal layer, with the curvature effects weaker, 
and that the RC correction fails to remedy the issue even over a significantly convex wall. The reason for the better response of the SST model, 
most probably driven by its $k$-$\omega$ branch, is unknown; curiously, the separation predictions are almost identical in both flows between 
DNS and models (with SA-RC) and there is no conjecture as to whether this reflects a fortuitous error cancellation, or the propagation of a kind of 
invariant boundary-layer quantity that is not degraded by the errors in skin friction.

\subsection{Evaluation of the RANS models in the flow field}

We sought a reason for the relative failure of the models by comparing what we view as revealing quantities between them and the DNS. 
These quantities include the primary shear stress normalized by the edge velocity, that is, $-\overline{u'v'}/U_\infty^2(1-C_p)$; 
the primary shear stress in local wall units, $-\overline{u'v'}/u_\tau^2$ (which is near 1 in very weak pressure gradient); 
and the parameter $a_{1,\mathrm{eff}} \equiv \mathcal{P}_k / S k$, where $\mathcal{P}_k = -\bar{u_i'u_j'} S_{ij}$ is the rate of production of turbulence kinetic energy $k$.  
The latter ratio can be viewed as a measure of the `efficiency' of the turbulence/mean energy-transfer process;   
it reduces to the Reynolds-stress `structure parameter' $a_1\equiv -\overline{u'v'}/k$, which has a nominal value near 0.3 in a two-dimensional shear layer.
We also considered the effective eddy viscosity $\mu_{t,\mathrm{eff}} = \rho \mathcal{P}_k / S^2$, based on the rate of turbulence-kinetic-energy production. 
This last quantity is most relevant, even for two-equation models, since it is very acceptable to have compensating errors that lead to the correct turbulent viscosity.
Unfortunately, no striking findings were made, but we display figures that illustrate the rich features of the solution.

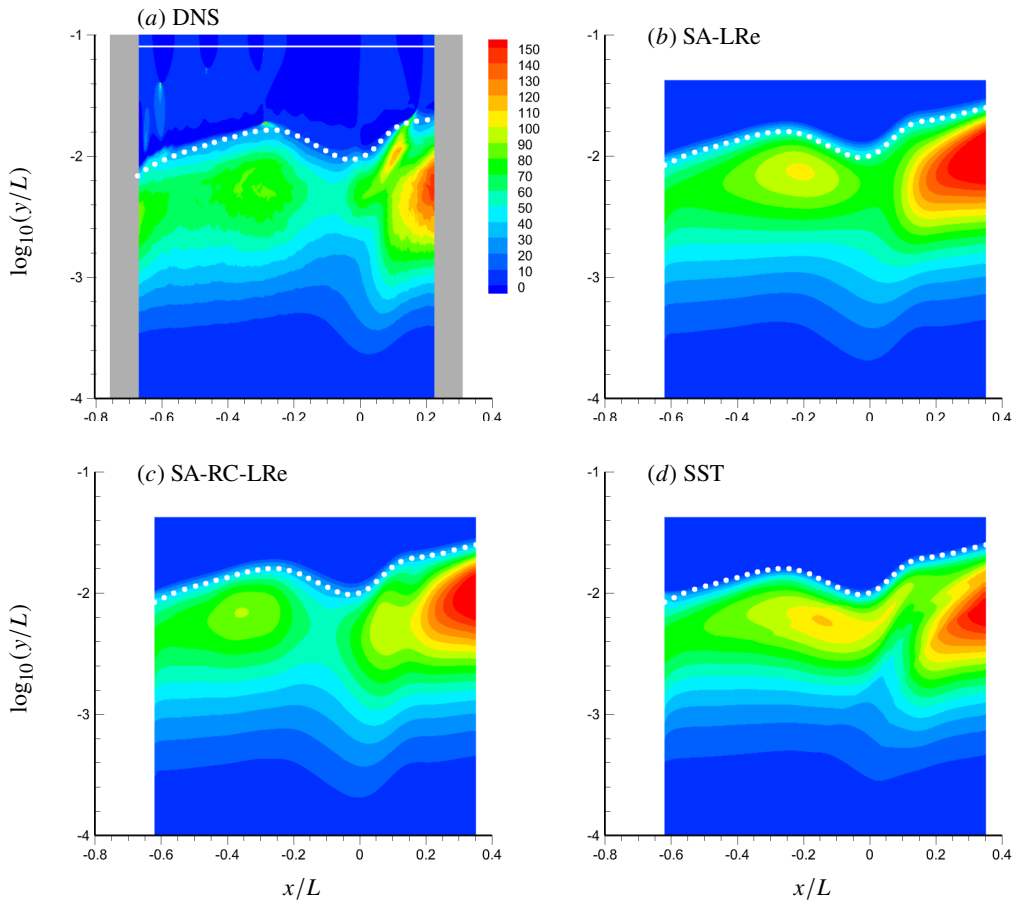
\begin{figure}
\centerline{
\hbox{
\rotatebox{90}{\put(063,000) {$\log_{10}(y/L)$}}
\includegraphics[trim={20 10 10 10},clip,width=0.50\textwidth]{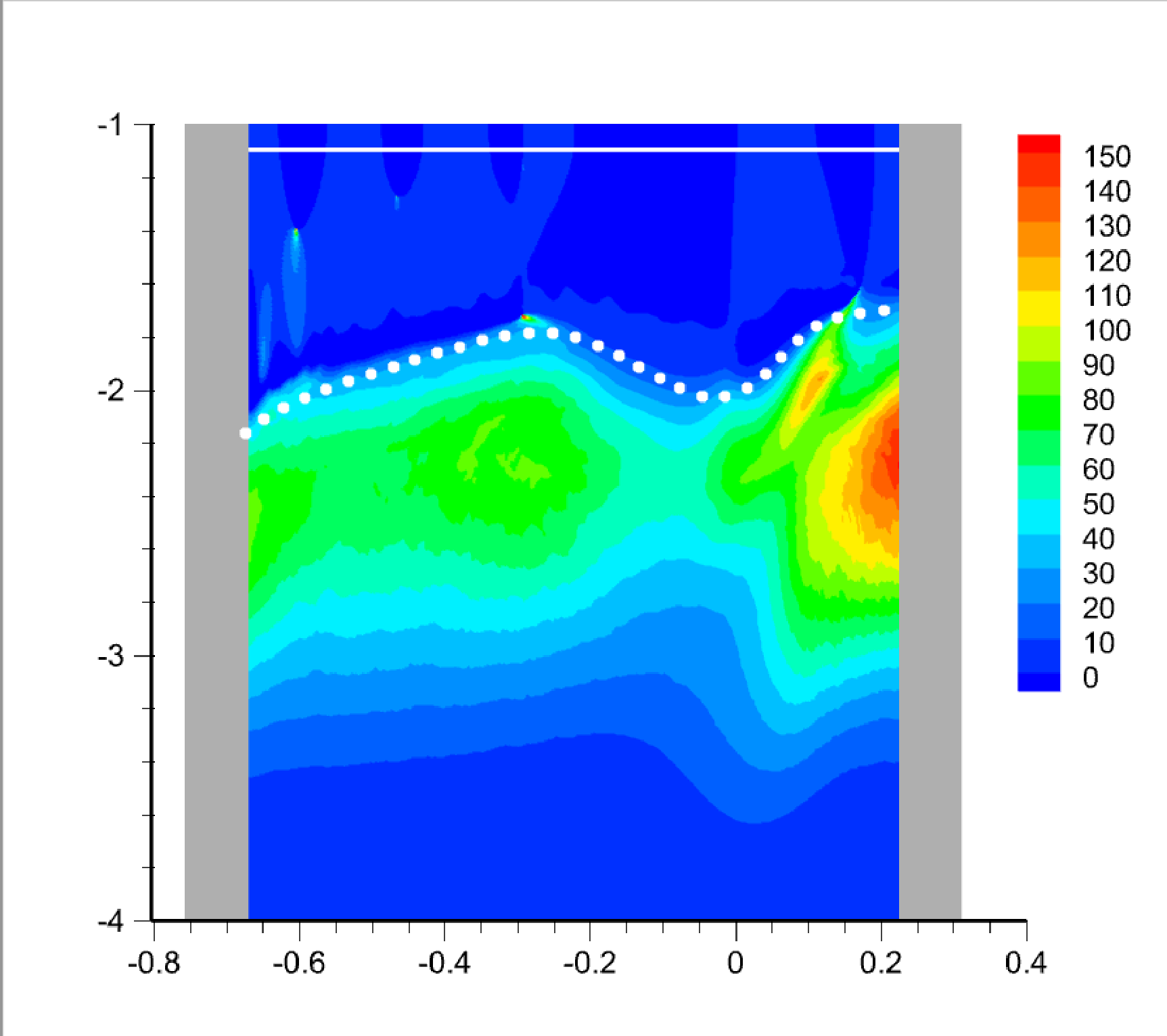}
\put(-155,157) {\c{black}{($a$) DNS}}
\includegraphics[trim={20 10 10 10},clip,width=0.50\textwidth]{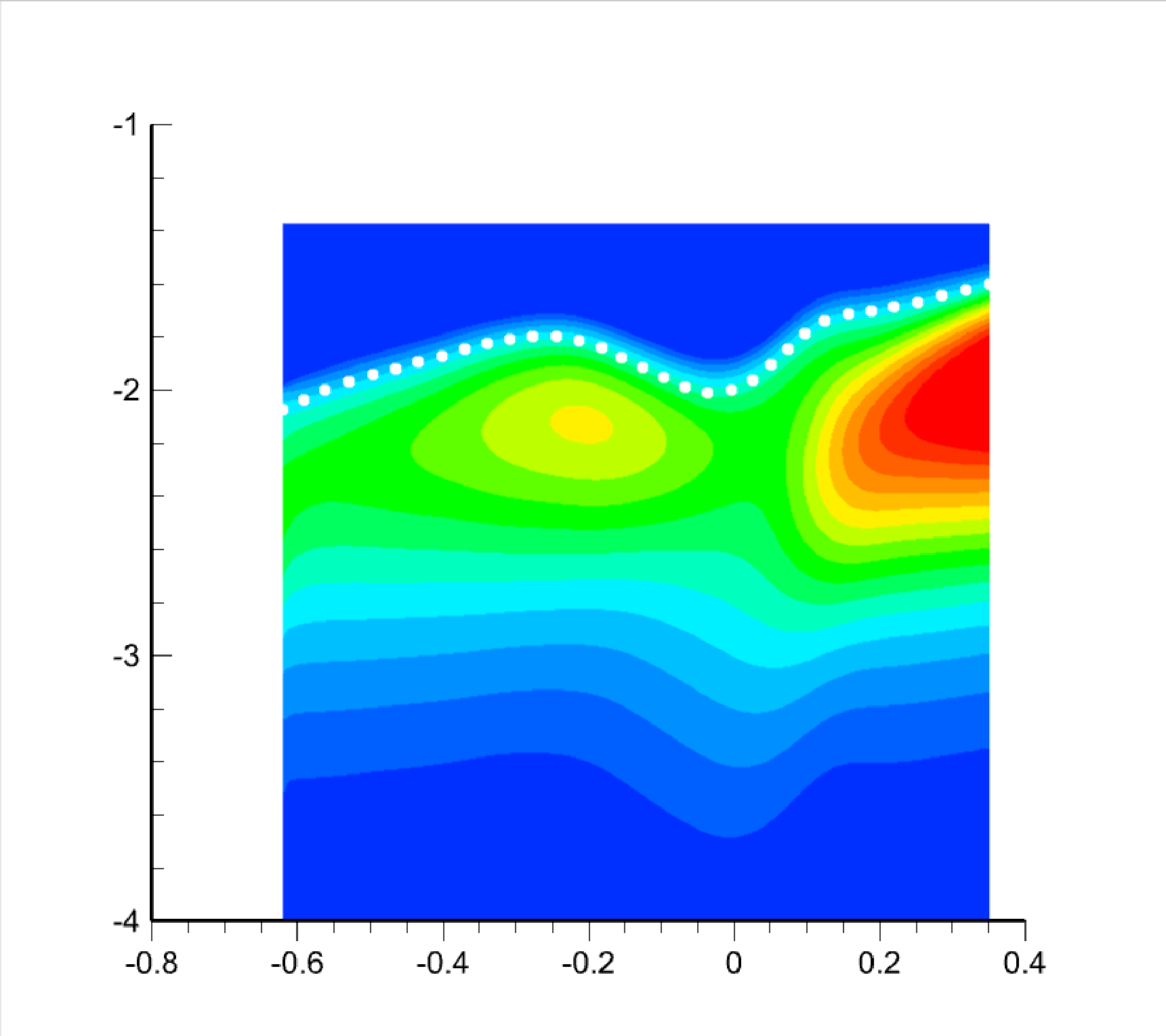}
\put(-155,150) {\c{black}{($b$)} SA-LRe}
}}
\vspace{-0.13truein}
\centerline{
\hbox{
\rotatebox{90}{\put(063,000) {$\log_{10}(y/L)$}}
\includegraphics[trim={20 10 10 10},clip,width=0.50\textwidth]{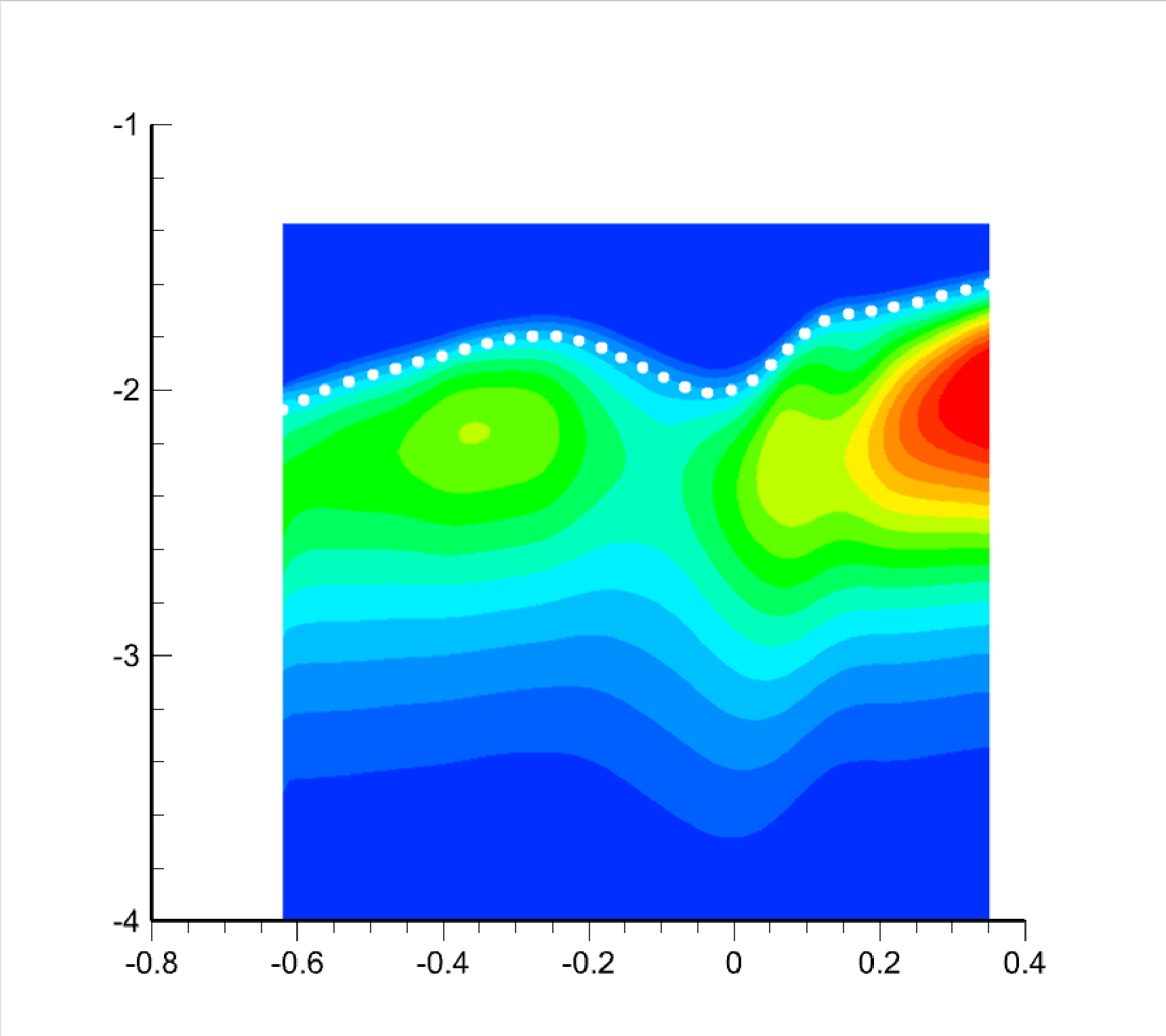}
\put(-155,150) {\c{black}{($c$)} SA-RC-LRe}
\put(-100,-005) {\c{black}{$x/L$}}
\includegraphics[trim={20 10 10 10},clip,width=0.50\textwidth]{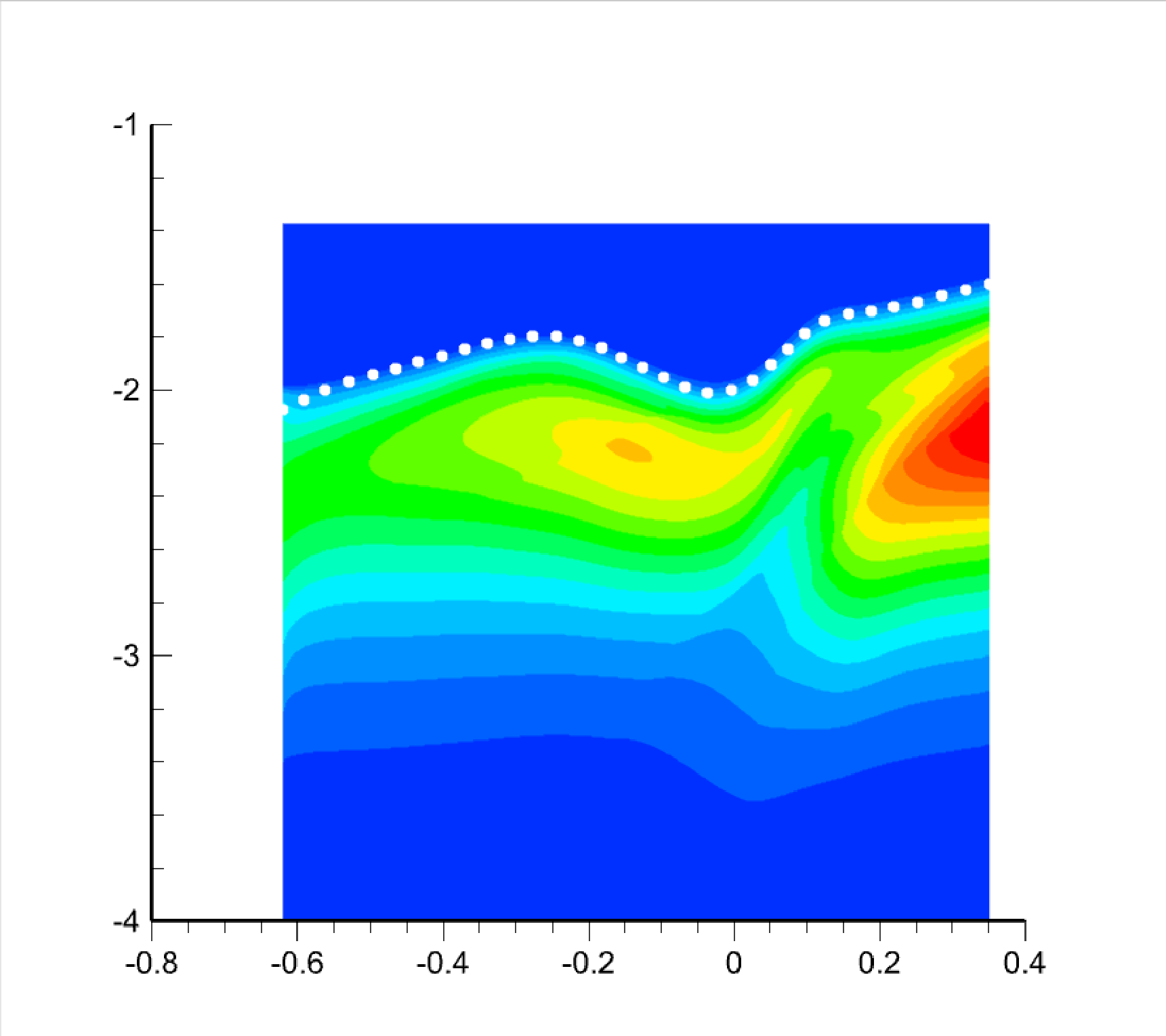}
\put(-155,150) {\c{black}{($d$)} SST}
\put(-100,-005) {\c{black}{$x/L$}}
}}
\vspace{0.20truein}
\caption{($a$) Effective and ($b$)--($d$) modelled eddy viscosity for Case~F: 
($a$)~DNS; 
($b$)~SA-LRe; 
($c$)~SA-RC-LRe; 
($d$)~SST. 
White symbols mark edge of boundary layer $\delta_{995}$, defined as in figure~\ref{fig:theta}.
In ($a$), grey zones are fringe regions, while white horizontal line denotes $y=Y_\mathrm{top}$. 
Contours are in units of molecular viscosity.  
}
\label{fig:nut(x,y)}
\end{figure}

Figure~\ref{fig:nut(x,y)} shows the eddy viscosity versus wall distance on a logarithmic axis, to \new{expose} the near-wall behavior. The edge of the boundary layer is marked, 
and $\delta$ is seen to drop by over 40\% in the FPG. The boundary-layer thickness is defined in terms of mean vorticity, 
as the location of the 99.5\%-of-edge-value of the `generalised velocity' \citep[see (3.1)--(3.2) of][]{CRS18}.
The DNS and the models each exhibit a saddle point, with that for the DNS (and SA-RC-LRe) near $x/L=-0.12$; 
the other quantities listed above also have a saddle point near this station in the DNS  -- {\em except\/} for the $a_1$ parameter, 
which has a single maximum, near $x=0$.  (This underlines the risk of relying too heavily on coordinate-dependent quantities; \cf figure~13$h$ of Coleman \etal 2018.) 

The eddy viscosity is contained within the vortical/boundary-layer region as expected; 
its edge is soft in the DNS, but sharper for the models. The rough inverted-parabola profile is present in the incoming layer; it then drops primarily
due to the thinning boundary layer, even though the edge velocity is rising.
We cannot assert that their downward evolution of the eddy viscosity is not a sign of partial relaminarization,
which could be cited as a reason for models to err. For instance, if $-\overline{u'v'}/u_\tau^2$ dropped much below 1/2, a description as `not 
truly turbulent' would come to mind for the sub-boundary layer.
As it is, this quantity only drops to about 0.55; thus the finding is a little ambiguous.

The comparison of effective/eddy viscosity is quite favorable to the RC correction, over the convex FPG region. 
The uncorrected SA and especially the SST model are less successful in capturing this feature.
At the station corresponding to the crest of the (virtual) bump, higher levels are seen very near the wall, 
driven by the high value of skin friction (normalized with $U_\infty^2$).
Beyond $x/L\approx 0.2$, the peak eddy viscosity grows rapidly as a result of the boundary layer rapidly thickening and allowing the length
scale of the turbulence to grow. 

The overall impression appears
justified that the models capture a wide variety of the features of the true turbulence field, which one would not expect for such minimal
descriptions, with at most two quantities. Recall that the SA model was calibrated only in zero pressure gradient, while the SST model depends 
on its $a_1$, which was guided by pressure-gradient cases, but still is only a single number. Another remark is that the present fields appear 
quite ready for Machine Learning.

\subsection{Attempt at improving the Spalart-Allmaras model} 

In both the C and F cases, the SA model generates too much skin friction in and following the FPG. In such a region, 
the skin friction is rapidly rising in the streamwise direction, and therefore the eddy viscosity also is, 
which is expressed by the relationship $\D\tilde{\nu}/\D t>0$. This led to the idea of an empirical reduction, 
specifically by multiplying the classical sum of terms of the SA model that produce $\D\tilde{\nu}/\D t$ by a function 
$f_{PG}\equiv 1-C\times f_w(r)$ where $C$ is a constant with a tentative value of 0.3.

The design principle for $f_{PG}$ was that it is neutral in free shear flows since it equals 1 there, and also in channel flows, where $\D\tilde{\nu}/\D t=0$. 
In flat-plate boundary layers, $\D\tilde{\nu}/\D t$ is very small. Therefore, the modified model is very backwards-compatible with the important validation cases. 
This motivated a simple test, with plans for a wider validation if it was successful.

The FUN3D solver was modified by Dr.~Li Wang of NASA Langley and tested on the F flow. The modification unfortunately made an insignificant difference. 
Our interpretation is that the sum of the terms entering $\D\tilde{\nu}/\D t$ is so much smaller than the individual terms of production, 
destruction and diffusion that the eddy viscosity remains essentially in equilibrium with the local skin friction in the sense that 
$\nu_t=\kappa d u_\tau$, thus defeating any modification that revolves around $\D\tilde{\nu}/\D t$. We believe that the brief description of a failed attempt 
is instructive, both to convey the fact that this particular one is not promising and to outline some principles of RANS-model modifications.

\section{\new{Closing remarks}}

The idea of comparing two boundary layers with the same non-uniform pressure distribution but with only the first 
one having wall curvature appears to have some merit. Calculating the transpiration distribution for the second 
flow is not trivial, and is possible only if the shared pressure distribution is very smooth and the transpiration 
line not too distant. It is not clear that an experimental duplication is possible; recall that \cite{SM73} had a 
constant pressure distribution which they obtained by adjusting both the curvature of the test wall and the 
position of the opposite wall. The pressure agreement achieved here is definitely close enough for convincing 
observations to be made up to the separation point. Concave and then convex curvature raises and reduces skin friction, respectively, and 
this with a delay. Transport turbulence models such as those we tested in principle can capture this, but they suffer 
from a wide inaccuracy irrespective of curvature as a result of the favorable pressure gradient alone. Deliberate 
modifications of models have yet to be attempted; the development of an internal layer roughly 10 times 
thinner than the full boundary layer is a very visible phenomenon and places this FPG region outside the known 
validation base of the models. The present quantitative results add to the Speed Bump set of data (although in a 
two-dimensional version) and therefore to the overall turbulence knowledge base.

\new{The major weakness of the present study is the lack of precise agreement of the mean/turbulence profiles 
at the inflow (\ie in the weak-APG region upstream of the strong FPG induced by the surface curvature) between the two DNS cases, and 
the ambiguity that introduces. This illustrates the non-trivial reality of matching incoming boundary layers when 
comparing RANS and/or scale-resolving simulations of the same flow.  There is much room for improvement here.} 

Future work along the present lines could involve a `negative bump' with convex-concave-convex and FPG-APG-FPG-APG
sequences but avoiding flow reversal near $x/L=0$ so that the pressure distribution is again transferable. Another possibility
would be a turning surface with a zero-convex-zero curvature sequence, emulating a deflected airplane control surface; airliner 
rudders are a prime example. Still, the most urgent task is to better understand the internal layer and the possibilities to train 
the turbulence models to address it.

\vfill

\backsection[Acknowledgements]{We are indebted to J.-R.~Carlson, C.\,L.~Rumsey, and L.~Wang for, respectively, 
creating the special boundary conditions, generating the grids, and making the source-code modifications 
needed for the FUN3D RANS studies described above.  P.\,S.~Iyer made useful comments on the manuscript. Computational resources were provided 
(for the RANS studies) by the NASA LaRC K-cluster and (for the DNS) the Pleiades cluster administered by the NASA Advanced Supercomputing (NAS) Division.  
G.\,N.\,C. is supported by the Transformational Tools and Technologies (TTT) Project of the Transformative Aeronautics Concepts Program 
under the NASA Aeronautics Research Mission Directorate.}

\backsection[Declaration of interests]{The authors report no conflict of interest.}

\backsection[Data availability statement]{The data that support the findings of this study are openly available on the 
NASA Turbulence Modeling Resource~(TMR) website: https://turbmodels.larc.nasa.gov.}





\bibliographystyle{our_jfm}

{}

\end{document}